\documentclass[journal]{IEEEtran}
\usepackage{multirow}
\usepackage{lineno}
\usepackage[nospace,noadjust]{cite}
\ifCLASSINFOpdf
\else
\usepackage[dvips]{graphicx}
\usepackage{color}
\fi
\usepackage[cmex10]{amsmath}
\usepackage{amssymb, amsmath}
\usepackage{amsmath}
\usepackage{graphicx}
\usepackage{multicol}
\usepackage{blindtext}
\usepackage{caption}

\begin{document}
\title{Ultrasonic Tissue Reflectivity Function Estimation Using Correlation Constrained Multichannel FLMS Algorithm with Missing RF Data 
\author{Jayanta~Dey and Md.~Kamrul~Hasan}
\thanks{M. K. Hasan and Jayanta Dey are with the Department
of Electrical and Electronic Engineering, Bangladesh University of
Engineering and Technology, Dhaka, Bangladesh (e-mail: khasan@eee.buet.ac.bd).}}
\maketitle

\begin{abstract}
Poor resolution of ultrasound images due to convolution of the tissue reflectivity function (TRF) with the system point spread function (PSF) is a major issue in medical ultrasound imaging. In this paper, we propose a correlation constrained missing-data estimation based blind multichannel frequency-domain least-mean-squares (\textit{md-b}MCFLMS) algorithm to undo the effect of PSF on the ultrasound radio-frequency (RF) data. In the first step, a block-based MCFLMS (\textit{b}MCFLMS) algorithm is proposed to estimate the TRFs and the PSF which are used in the second step to estimate the missing data. This missing data is used in the \textit{md-b}MCFLMS algorithm to construct a modified cost function for further improvement of the image resolution. To account for the nonstationarity of the PSF, unlike the blocking approach described in the literature, we introduce a time-efficient blocking method in this paper. The blocking approach described here uses a block position independent fixed size matrix and can be implemented parallely. The  \textit{b}MCFLMS algorithm, however, shows misconvergence due to both channel noise and propagation of TRF estimation error from the previous blocks. This phenomenon is more intense in the case of \textit{md-b}MCFLMS algorithm because of increased estimation error. To address this problem, a novel constraint based on the correlation between the measured RF data and estimated TRF is proposed in this paper. The efficacy of our proposed blind deconvolution algorithm is measured using simulation phantom, experimental phantom and \emph{in-vivo} data. The proposed \textit{md-b}MCFLMS algorithm shows normalized projection misalignment (NPM) improvement of about $2\sim6.98$ dB and resolution gain (RG) improvement of $0.58\sim6.06$ dB compared to other techniques  in the literature. Moreover, because of the frequency-domain implementation it is computationally more efficient, fast converging and robust than its time-domain counterpart \textit{$l_1$-b}MCLMS algorithm reported in the literature.


\end{abstract}

\section{INTRODUCTION}
\IEEEPARstart{U}{ltrasound} imaging is non-invasive and non-ionizing in nature which have made it popular in medical diagnostics \cite{jensen1993deconvolution}. It has added benefits of real-time imaging, low cost and portability compared to X-ray, magnetic resonance imaging and other imaging modalities. However, clinical ultrasound images are often difficult to interpret due to blurring, speckle noise and low contrast between different soft tissues \cite{shin2009sensitivity}. Point spread function (PSF) of ultrasound imaging system introduces blurring and thereby degrades the resolution of ultrasonic images. The removal of PSF effect from the ultrasound images can restore the resolution of images and thus improve the diagnostic quality of ultrasound imaging.

Use of a suitable model for the measured ultrasound RF data is the key to the development of an effective deconvolution technique for the ultrasound images. In ultrasound imaging, the transmitted ultrasonic pulse reflected by the randomly distributed scatterers present in the tissue is received by a transducer array to generate the diagnostic image. Considering the scattering of the pulse in the tissue as weak, we can use the first order Born approximation and consider the tissue scattering system as a linear system. Therefore, we can model the received RF signal as the  convolution between the PSF and the transfer function of the scattering system, i.e, TRF \cite{jensen1991model}. In this process blurring is introduced in the RF data by the PSF.  A number of algorithms has been proposed in the literature to deconvolve the PSF from the received RF data and thereby restore the image resolution. These algorithms can be classified into two groups. The first group estimates the PSF first and then a classical deconvolution algorithm is applied to estimate the TRFs. These methods are based on homomorphic filtering \cite{taxt1995restoration}, \cite{michailovich2004phase} which involves filtering out the PSF either in the cepstrum domain \cite{taxt1995restoration}, \cite{michailovich2007blind} or in the log magnitude domain \cite{adam2002blind}. These algorithms are elegant in the sense that they are simple and can be implemented in real-time. Filtering the wavelet coefficients of the log magnitude spectrum gives better result in terms of mean square error (MSE) than the cepstrum based methods  \cite{michailovich2007blind}. However, tuning the length of the filter in cepstrum domain and selecting the decomposition level of wavlet decomposition determines the smoothness of the estimated PSF \cite{michailovich2002shift} and hence the overall deconvolution accuracy. In addition, this category of techniques assume that the spectrum of the PSF and the TRF lie in separate spectral band which is not completely true and again the minimum phase assumption for PSF that is applied here, may not be satisfied in general \cite{taxt1995restoration}. Moreover, inaccuracy in phase unwrapping poses another problem for these algorithms \cite{michailovich2007blind}. To solve the problem of phase unwarping, a recent hybrid parametric inverse filtering (HYPIF) algorithm \cite{michailovich2007blind} has been proposed which estimates the PSF in two steps. First, partial information of the PSF, i.e., power spectrum is estimated using the homomorphic filtering. This partial information obtained is used to constraint the shape of the inverse filter. Then linearity of the inverse filter is exploited to recover the phase of the PSF. However, in this method the energy of the inverse filter should be regularized to avoid instability where the PSF has zero or very low magnitude. Again finite number of coefficients using the Fourier basis cannot properly represent the finite support of the PSF. To solve this problem the Fourier basis is replaced by complex exponentials with compact support \cite{michailovich2007blind}. Nevertheless, they may not properly represent the PSF.

The second group of algorithms estimates the PSF and the TRFs simultaneously. Among them the blind deconvolution method described in \cite{yu2012blind} improves the convergence speed and reduces the computational load by projecting the TRF into the null space of the correlation matrix and the PSF onto the space spanned by the third-order B-spline wavlet basis. However, in the presence of noise thresholding is necessary to determine the null space bases. Another classical algorithm of iterative blind deconvolution \cite{ayers1988iterative} suffers from poor convergence property. The parametric methods (\cite{tekalp1990maximum}-\cite{reeves1992blur}) of this group depend on how accurately the parameters represent the ultrasound imaging system. In addition, due to extremely complex composition of most biological tissues, derivation of a convenient and accurate model for \emph{in-vivo} PSF is not possible \cite{michailovich2005novel}.

Lately, a block-based blind deconvolution method in the time-domain using the multichannel LMS (MCLMS) algorithm has been presented in \cite{hasan2016blind} which estimates the TRFs block by block to account for the nonstationarity of the PSF. The blocking procedure described in the time-domain method \cite{hasan2016blind} uses convolution matrix segmentation and thereby FFT is not applicable for convolution. Therefore, the algorithm is computationally inefficient due to the implementation of time-domain convolution using matrix multiplication. In addition, the size of the matrix increases as the number of blocks increases. Furthermore, it suffers from misconvergence in the presence of additive white noise and propagation of estimation error from block to block. An attempt was made to solve this problem using a damped variable step-size \cite{hasan2006damped}, gradient averaging and \textit{$l_1$}-norm constraint. Here in order to apply the \textit{$l_1$}-norm constraint, the data is assumed to be sparse which is not generally true for the case of \emph{in-vivo} data. The noise effect minimizing methods described here are adopted on ad hoc basis and the misconvergence may not be completely stopped even after applying all the aforesaid noise effect minimization techniques.

In this paper, we propose a missing data estimation based blind deconvolution algorithm in the frequency-domain with correlation constraint for noise corrupted ultrasound RF data. Due to smaller eigenvalue spread, a frequency-domain approach facilitates faster convergence of the adaptive algorithms compared to the time-domain ones \cite{beaufays1995transform}, \cite{ahmad2006extended}. Moreover, the variable step-size (VSS) MCFLMS algorithm is known to be noise robust \cite{haque2007variable}, \cite{gaubitch2006generalized}. Unlike the blocking procedure described in \cite{hasan2016blind}, in this paper we introduce a new blocking technique that facilitates the use of FFT and make the algorithm computationally efficient. However, as with the other reported crossrelation based blind adaptive algorithms \cite{haque2008noise}, \cite{liao2015analysis}, the proposed algorithm also suffers from misconvergence in the presence of channel noise and estimation error from the previous estimated blocks of the TRFs. To overcome this problem, we propose a novel noise effect compensating constraint based on the correlation between the measured RF data and estimated TRFs that can compensate the effect of noise and estimation error stated above. Next, for estimating the missing RF data, the PSF is estimated using the R-MINT algorithm in the same manner described in \cite{hasan2016blind}. Estimated PSF and TRFs are used to estimate the missing data which were not available due to the nature of data acquisition of the ultrasound imaging. This estimated missing data is then used to develop the \textit{md-b}MCFLMS algorithm to improve the resolution of the ultrasound images further. The performance of the proposed blind deconvolution techniques is evaluated on the simulation phantom, experimental phantom and \emph{in-vivo} data.

The paper is organized as follows. Section II describes the SIMO model of an ultrasound imaging system. The block-based blind deconvolution algorithm and the noise effect compensating constraint are explained in section III. Further improvement of resolution of ultrasound images using the estimated missing RF data is discussed in section IV. The performance of \textit{md-b}MCFLMS algorithm is demonstrated in section V. Finally, summarizing the contributions with highlights for future research the paper concludes in section VI.

%%%%%%%%%%%%%%%%%%%%%%%%%%%%%%%%%%%%%%%%%%%%%%%%%%%%%%%%%%%%%%%%%%%%%%%%%%
\section{Problem Formulation}
In the ultrasound imaging system, an array of piezoelectric elements sequentially emit the same ultrasound pulse in the tissue and receive echo signals from multiple A-lines. 
\begin{figure}[h]
\includegraphics[scale=.6]{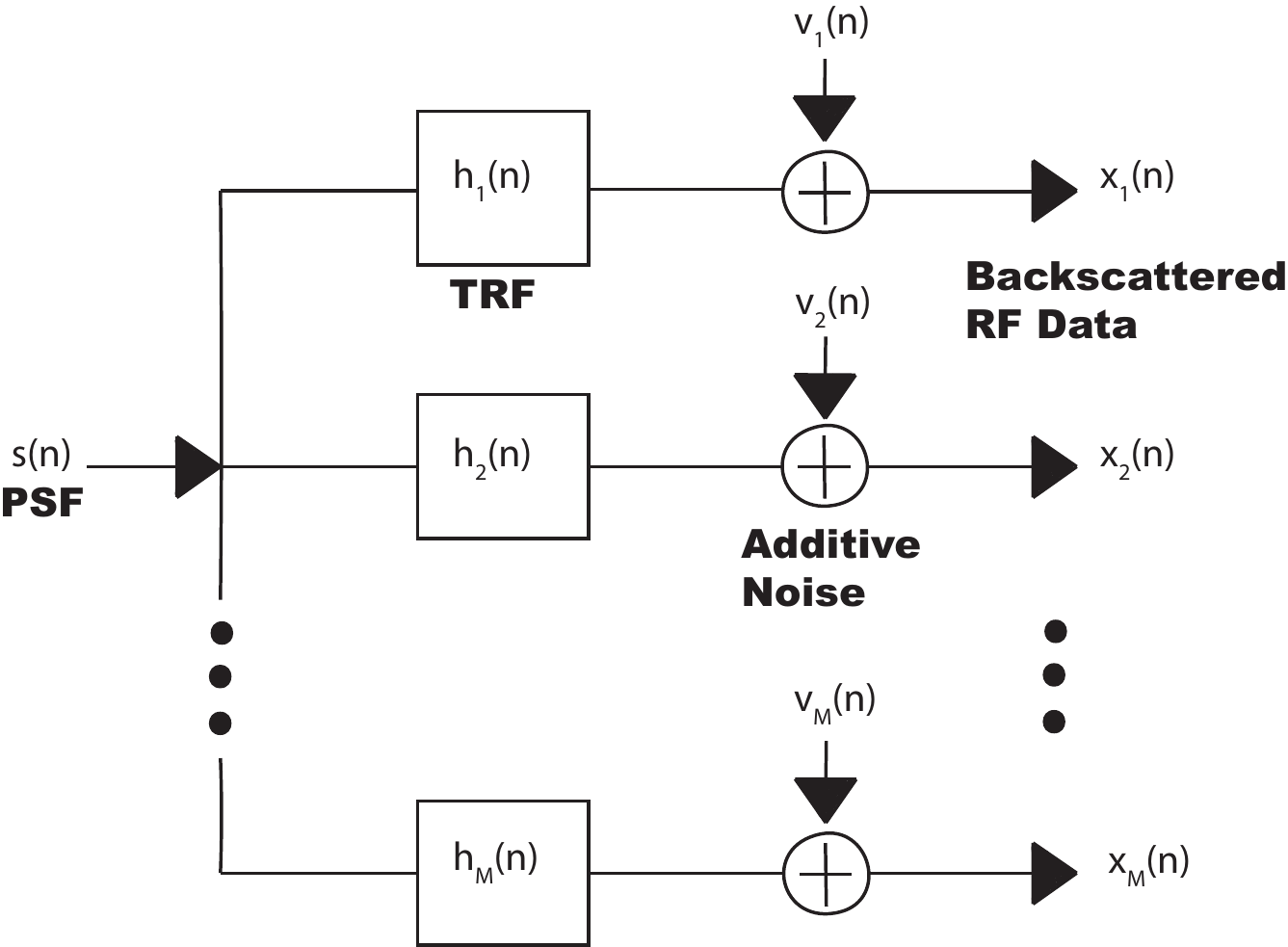}
\caption{Illustration of ultrasonic data acquisition system as SIMO model, showing the relationship between the backscattered RF data $x_i(n)$ and the point spread function, $s(n)$.}
\label{fig:simo}
\end{figure}
This system can be modeled as a SIMO model with the ultrasound pulse as the single input, the measured echo signal lines as multiple outputs, and the TRFs along the axial direction as multiple system channels \cite{yu2012blind}. Figure \ref{fig:simo} shows the SIMO model of the backscattered ultrasound RF data, where the ultrasound pulse or the PSF $s(n)$ convolves with the $i$-th channel transfer function or TRF $h_i(n)$. With additive noise $v_i(n)$, the measured RF data are given by

\begin{equation}\label{eqn:model}
x_i(n) = s(n)*h_i(n) + v_i(n), i = 1,2,\cdots,M
\end{equation}
where $x_i(n)$ denotes the backscattered RF data of the $i$-th scan line and $M$ is the number of total scan lines. If the length of $x_i(n)$ is $L$ and that of $s(n)$ is $L_s$ with $L_s<<L$, we can write \eqref{eqn:model} in matrix form as

\begin{equation}
\mathbf{x}_i(n) = \mathbf{S}(n)\mathbf{h}_i + \mathbf{ v}_i(n)
\end{equation}
where $\mathbf{S}(n)$ is the $(L + L_s - 1)\times L$ convolution matrix constructed from $s(n)$, and
\begin{align*}
\mathbf{h}_i &= 
\begin{bmatrix}
h_i(n) &h_i(n-1) \cdots &h_i(n-L+1)
\end{bmatrix}^T\\
\mathbf{x}_i(n) &= 
\begin{bmatrix}
x_i(n) &x_i(n-1) \cdots &x_i(n-L-L_s+2)
\end{bmatrix}^T\\
\mathbf{v}_i(n) &= 
\begin{bmatrix}
v_i(n) &v_i(n-1) \cdots &v_i(n-L-L_s+2)
\end{bmatrix}^T
\end{align*}

However, as each data sample of the ultrasound RF signal results from the reflection of the ultrasound pulse from a scatterer of the tissue, in reality, we have RF echo data equal to the TRF length $L$ instead of $L+L_s-1$. Moreover, \eqref{eqn:model} assumes that the PSF is stationary, i.e., remains constant while penetrating the tissue. But, if the acquired RF data is long in the axial direction, the PSF suffers from the depth dependent attenuation while traveling through the tissue. Then this nonstationary PSF restricts the direct use of the crossrelation based MCFLMS algorithm for blind SIMO model identification in the deconvolution of ultrasound images \cite{hasan2016blind}. Therefore, the main challenge in ultrasound deconvolution process is to estimate the TRFs, $h_i(n), i=1,2,\cdots,M$, using the truncated and nonstationary RF data corrupted by additive white noise. Here the deconvolution process is carried out in the frequency-domain to obtain more robustness to noise than in the time-domain along with faster convergence facilitated by smaller eigenvalue spread and reduced computational complexity.

%To be identifiable, the TRFs should fulfill the following two conditions:
%\begin{enumerate}
%\item[1.] The TRFs do not contain any common zero
%\item[2.] The autocorrelation matrix for the RF data should be of full rank
%\end{enumerate}
%Fortunately, both the conditions are satisfied in ultrasound imaging.

%%%%%%%%%%%%%%%%%%%%%%%%%%%%%%%%%%%%%%%%%%%%%%%%%%%%%%%%%%%%%%%%%%%%%%%%%%%%%%
\section{\bf{Method}}
\subsection{Fundamentals and Frequency Domain Approach}
For noise-free case, the crossrelation error defined in the following can be exploited to estimate the TRFs:
\begin{align}\label{eqn:full_err}
e_{ij}(n) = x_i(n)*h_j(n) - x_j(n)*h_i(n),~~ i,j &= 1, 2, \cdots, M, \nonumber \\
 						i &\neq j
\end{align}
Note that the error function in \eqref{eqn:full_err} becomes zero when  $x_i(n)$ in \eqref{eqn:model} for noiseless condition is substituted into it. However, as shown in \cite{hasan2016blind}, due to incomplete data acquisition in ultrasound imaging only the first $L$ samples of the error function will be equal to zero. Taking this into account, the truncated error function in matrix form is written as
\begin{equation}\label{eqn:truncated_error}
\tilde{\mathbf{e}}_{ij} = \mathbf{D}(\mathbf{C}_{\tilde{x}_i}\mathbf{h}_j - \mathbf{C}_{\tilde{x}_j}\mathbf{h}_i)
\end{equation}
where, 
\begin{equation}
\tilde{x}_i(n) = x_i(n), n = 0, 1, 2, \cdots, L-1, \nonumber
\end{equation}
$\mathbf{C}_{\tilde{x}_i}$ is the convolution matrix formed with the truncated RF data $\tilde{x}_i(n)$ of length $L$ and
\begin{align*}
 \mathbf{D} = 
\begin{bmatrix}
\mathbf{I}_{L\times L}   & \mathbf{0}_{L\times (L-1)}
\end{bmatrix}
\end{align*}
is the truncation matrix, $\mathbf{I}$ is the identity matrix of size $L\times L$, $\mathbf{0}$ is the null matrix of size $L\times (L-1)$, 
\begin{align*}
\mathbf{h}_i &= 
\begin{bmatrix}
h_i(n) &h_i(n-1) \cdots &h_i(n-L+1)
\end{bmatrix}^T
\end{align*}
 
Due to nonstationarity of the PSF in ultrasound imaging, \eqref{eqn:truncated_error} cannot be directly used to estimate the TRFs. An appropriate solution to this problem is to estimate the TRFs block by block from the error blocks formed from \eqref{eqn:truncated_error} as shown in \cite{hasan2016blind}. Then we can consider the PSF as quasi-stationary within that particular block. However, the blocking approach described in \cite{hasan2016blind} requires matrix multiplication between segments of a convolution matrix formed from the RF data and estimated TRF. Its frequency-domain implementation will again require multiplication of an error block with a DFT matrix making the algorithm computationally inefficient. Unlike in \cite{hasan2016blind}, here we propose a different blocking approach that uses smaller and fixed size vector operations with FFT applicability and thus a faster approach.

The crossrelation error in \eqref{eqn:full_err} is basically the difference between two convolution operations. Therefore, if we can implement a full convolution as a summation of convolutions between smaller blocks of signals, \eqref{eqn:truncated_error} can be implemented efficiently by eliminating the need of using the truncation operator $\mathbf{D}$ as well as the unnecessary computation of the full convolution (the bracketed part in \eqref{eqn:truncated_error}). Before we go into details, we first show how a convolution can be implemented block by block. In the subsequent discussion, ` $\tilde{}$ ' is used to denote truncated data, channel number is placed as subscript and block number is presented in the superscript. Now consider the convolution between the trucated RF data $\tilde{x}_i(n)$ of an arbitrary channel $i$ and the estimated TRF data $ \hat{h}_j(n)$ of the $j$-th channel given by
\begin{equation}\label{eqn:convl}
y_{ij}(n) = \tilde{x}_i(n)*h_j(n)
\end{equation}
The $z$-transform of $h_j(n)$, denoted as $H_j(z)$ can be expressed as
\begin{align}
H_j(z) &= h_j(0) + h_j(1) z^{-1} + \cdots + h_j(L_b-1) z^{-(L_b-1)}\nonumber \\
&+ h_j(L_b) z^{-L_b} + \cdots + h_j(BL_b)z^{-(BL_b-1)}\nonumber\\
	&= H_j^1(z) + z^{-L_b} H_j^2(z) + \cdots + z^{-(B-1)L_b} H_j^B(z)
\end{align}
where $H_j^b(z)$ can be written as
\begin{align}\label{eqn:H}
H^b_j(z) &= h_j\big((b-1)L_b) + h_j\big((b-1)L_b + 1)z^{-1} \nonumber\\
	&~~~+ \cdots + h_j\big((bL_b-1))z^{-(L_b-1)},b = 1,2,\cdots, B
\end{align}
Here $B$ is the total number of blocks and $L_b = floor(L/B)$.\\
Similarly, for the $i$-th channel RF data $x_i(n)$, we can write
\begin{equation}\label{eqn:X}
\tilde{X}_i(z) = \tilde{X}_i^1(z) + z^{-L_b} \tilde{X}_i^2(z) + \cdots + z^{-(B-1)L_b} \tilde{X}_i^B(z)
\end{equation}
where 
\begin{align}
\tilde{X}^b_i(z) &= \tilde{x}_i\big((b-1)L_b) + \tilde{x}_i\big((b-1)L_b + 1)z^{-1} \nonumber\\
	&~~~+ \cdots + \tilde{x}_i\big((bL_b-1))z^{-(L_b-1)},b = 1,2,\cdots
\end{align}
Using \eqref{eqn:H} and \eqref{eqn:X} , \eqref{eqn:convl} can be written in the $z$-domain as
\begin{align}\label{eqn:z_trns}
Y_{ij}(z) &= \tilde{X}_i(z)H_j(z) \nonumber \\
          &= \tilde{X}_i^1(z)H_j^1(z) + z^{-L_b}\tilde{X}_i^1(z)H_j^2(z) + \nonumber \\ 
          ~& z^{-L_b}\tilde{X}_i^2(z)H_j^1(z) + z^{-2L_b}\tilde{X}_i^2(z)H_j^2(z) \nonumber \\ 
 		 ~&+ \cdots + z^{-2(B-1)L_b}\tilde{X}_i^B(z)H_j^B(z)
\end{align}
Taking the inverse $z$-transform of \eqref{eqn:z_trns}, we get
\begin{align}\label{eqn:block_conv}
y_{ij}(n) &=  \tilde{x}_i^1(n)*h_j^1(n) + z^{-L_b}\tilde{x}_i^1(n)*h_j^2(n) \nonumber\\
& + z^{-L_b}\tilde{x}_i^2(n)*h_j^1(n) + z^{-2L_b}\tilde{x}_i^2(n)*h_j^2(n) \nonumber \\
& + \cdots + z^{-2(B-1)L_b}\tilde{x}_i^B(n)*h_j^B(n)
\end{align}
Here multiplication by $z^{-1}$ refers to unit sample delay, and $\tilde{x}_i^b(n)$ and $h_j^b(n)$ represent the $b$-th block of the $i$-th channel RF and $j$-th channel TRF, respectively. As evident from \eqref{eqn:block_conv}, a convolution operation can be splitted into a sum of smaller convolution blocks of identical length. For reasons explained in \eqref{eqn:truncated_error}, we will consider the first $L$ samples of total $2L-1$ samples of the convolution in \eqref{eqn:block_conv}. Now modifying \eqref{eqn:block_conv} we can write:
\begin{align}\label{eqn:block_conv1}
\tilde{y}_{ij}(n) &= \tilde{x}_i^1(n)*h_j^1(n) + z^{-L_b}\tilde{x}_i^1(n)*h_j^2(n) \nonumber\\
&~~~+ z^{-L_b}\tilde{x}_i^2(n)*h_j^1(n) + z^{-2L_b}\tilde{x}_i^2(n)*h_j^2(n) + \nonumber \\
&~~~\cdots + z^{(B-1)L_b}\tilde{x}_i^B(n)h_j^1(n) + z^{(B-1)L_b}\tilde{x}_i^1(n)h_j^B(n) \nonumber\\
&= y_{ij}^{11}(n) + z^{-L_b}y_{ij}^{21}(n) + z^{-L_b}y_{ij}^{12}(n) + \cdots \nonumber\\
&~~~+ z^{(B-1)L_b}y_{ij}^{1B}(n) + z^{(B-1)L_b}y_{ij}^{B1}(n)
\end{align}
where
\begin{align*}
 y_{ij}^{pq}(n) = \tilde{x}_i^p(n)*h_j^q(n)
\end{align*}
\noindent
To account for the nonstationarity problem of ultrasound PSF, the convolution must be evaluated in smaller blocks of $L_b$ samples. Here each of the smaller convolutions in \eqref{eqn:block_conv1} is of length $2L_b-1$. Now if we observe \eqref{eqn:block_conv1}, it is evident that only the first $L_b$ samples of $y_{ij}^{11}(n)$ contributes to the first convolution block $\tilde{y}_{ij}^1$ and its last $L_b - 1$ samples contributes to the next convolution block $\tilde{y}_{ij}^2$. Adding to this the first $L_b$ samples of $y_{ij}^{12}(n)$ and $y_{ij}^{21}(n)$, we get the second convolution block $\tilde{y}_{ij}^2$ and so on.
As the first block does not represent the general idea behind the blocking technique, we explain the mathematical operations on the second convolution block of length $L_b$. Here two truncation matrices $\mathbf{A}_1$ and $\mathbf{A}_2$ are used to take the last $L_b-1$ and the first $L_b$ samples of a convolution, respectively. Now,
\begin{enumerate}
\item[1.] 
$\tilde{\mathbf{y}}_{ij}^{11} = \mathbf{A}_1\mathbf{C}_{\tilde{x}_i^1}\mathbf{h}_j^1 = \mathbf{A}_1\mathbf{y}_{ij}^{11}$\\
where $ \mathbf{A}_1 = 
\begin{bmatrix}
\mathbf{0}_{(L_b-1)\times L_b}   & \mathbf{I}_{(L_b-1)\times (L_b-1)}\\
\mathbf{0}_{1\times L_b}        & \mathbf{0}_{1\times (L_b-1)}
\end{bmatrix}
$\\

\item[2.]  
$\tilde{\mathbf{y}}_{ij}^{12} = \mathbf{A}_2\mathbf{C}_{\tilde{x}_i^1}\mathbf{h}_j^2 = \mathbf{A}_2\mathbf{y}_{ij}^{12}$\\
where $ \mathbf{A}_2 = 
\begin{bmatrix}
\mathbf{I}_{L_b\times L_b}   & \mathbf{0}_{L_b\times (L_b-1)}\\
\end{bmatrix}
$\\

\item[3.]  
$\tilde{\mathbf{y}}_{ij}^{21} = \mathbf{A}_2\mathbf{C}_{\tilde{x}_i^2}\mathbf{h}_j^1 = \mathbf{A}_2\mathbf{y}_{ij}^{21}$\\
\end{enumerate}
Therefore, the second block of $\tilde{\mathbf{y}}_{ij}$ described in \eqref{eqn:block_conv1} is given by
\begin{align*}
\tilde{\mathbf{y}}_{ij}^2 &= \tilde{\mathbf{y}}_{ij}^{11} + \tilde{\mathbf{y}}_{ij}^{12} + \tilde{\mathbf{y}}_{ij}^{21}\\
					    &= \mathbf{A}_1\mathbf{y}_{ij}^{11} + \mathbf{A}_2\mathbf{y}_{ij}^{12} + \mathbf{A}_2\mathbf{y}_{ij}^{21}\\
					    &= \sum_{p=1}^{1}\mathbf{A}_1\mathbf{y}^{p(2-p)}_{ij} + \sum_{p=1}^{2}\mathbf{A}_2\mathbf{y}^{p(2-p+1)}_{ij}
\end{align*}
In general, for any block $b$, the last $L_b-1$ length of the following convolutions contribute to the $b$-th block convolution $\tilde{y}_{ij}^b$:
\begin{equation}\label{eqn:last_conv}
\mathbf{y}_{ij}^{p(b-p)} = \mathbf{C}_{\tilde{x}_i^p}\mathbf{h}_j^{(b-p)} ,
p = 1,2,\cdots,b-1
\end{equation}
And the first $L_b$ length of the following convolutions contribute to $\tilde{y}_{ij}^b$:
\begin{equation}\label{eqn:first_conv}
\mathbf{y}_{ij}^{p(b-p+1)} = \mathbf{C}_{\tilde{x}_i^p}\mathbf{h}_j^{(b-p+1)},
p = 1,2,\cdots,b 
\end{equation}
The convolution between $\tilde{x}_i(n)$ and $h_j(n)$  for the $b$-th block is then given by
\begin{equation}\label{eqn:blocked_convolution}
\tilde{\mathbf{y}}_{ij}^b = \sum_{p=1}^{b-1}\mathbf{A}_1\mathbf{y}^{p(b-p)}_{ij} + \sum_{p=1}^{b}\mathbf{A}_2\mathbf{y}^{p(b-p+1)}_{ij}
\end{equation}

\subsection{\textbf{b}MCFLMS Algorithm for TRF estimation}
As the truncated crossrelation error described in \eqref{eqn:truncated_error} is the difference between two convolutions, a similar approach as described in \eqref{eqn:last_conv}, \eqref{eqn:first_conv} and \eqref{eqn:blocked_convolution} can be adopted to evaluate the crossrelation error block by block. In a similar way to \eqref{eqn:last_conv}, the last $L_b-1$ samples of the following error function contributes to the $b$-th block crossrelation error:
\begin{equation}\label{eqn:last_err}
\mathbf{e}_{ij}^{p(b-p)} = \mathbf{C}_{\tilde{x}_i^p}\hat{\mathbf{h}}_j^{(b-p)} - \mathbf{C}_{\tilde{x}_j^p}\hat{\mathbf{h}}_i^{(b-p)} ,
p = 1,2,\cdots,b-1
\end{equation}
where $\mathbf{e}_{ij}^{p(b-p)}$ is the crossrelation error considering the $p$-th block of RF data and $(b-p)$-th block of TRF data. Notice that now the true TRF has been replaced by the estimated TRF. In the same manner, according to \eqref{eqn:first_conv}, the first $L_b$ samples of the following error function contributes to the $b$-th block crossrelation error:
\begin{equation}\label{eqn:first_err}
\mathbf{e}_{ij}^{p(b-p+1)} = \mathbf{C}_{\tilde{x}_i^p}\hat{\mathbf{h}}_j^{(b-p+1)} - \mathbf{C}_{\tilde{x}_j^p}\hat{\mathbf{h}}_i^{(b-p+1)},
p = 1,2,\cdots,b 
\end{equation}
Therefore, the crossrelation error for the $b$-th block is
\begin{align}\label{eqn:error}
\tilde{\mathbf{e}}_{ij}^b &= \sum_{p=1}^{b-1}\mathbf{A}_1\mathbf{e}^{p(b-p)}_{ij} + \sum_{p=1}^{b}\mathbf{A}_2\mathbf{e}^{p(b-p+1)}_{ij} \nonumber\\
			 &= \sum_{p=1}^{b-1} \mathbf{A}_1\mathbf{e}^{p(b-p)}_{ij} + \sum_{p=2}^{b}\mathbf{A}_2\mathbf{e}^{p(b-p+1)}_{ij} + \mathbf{A}_2\mathbf{e}^{1b}_{ij}\nonumber\\
			 &= \sum_{p=1}^{b-1} \tilde{\mathbf{e}}^{p(b-p)}_{ij} + \sum_{p=2}^{b}\tilde{\mathbf{e}}^{p(b-p+1)}_{ij} + \tilde{\mathbf{e}}^{1b}_{ij}
\end{align} 
Notice that the error components $\tilde{\mathbf{e}}^{p(b-p)}_{ij}$ and $\tilde{\mathbf{e}}^{p(b-p+1)}_{ij}$ that constitute the $b$-th block error function $\tilde{\mathbf{e}}_{ij}^b$, can be computed parallely. Here only the third term of the right side of \eqref{eqn:error} depends on the $b$-th block TRF, $\hat{\mathbf{h}}^b$ and the first two terms depend on $\hat{\mathbf{h}}^1,\hat{\mathbf{h}}^2,\cdots,$ and $\hat{\mathbf{h}}^{b-1}$. While estimating the TRF of the $b$-th block, all the TRF $\hat{\mathbf{h}}^q$ for $q = 1,2,\cdots,b-1$ are already known and therefore, may be treated as constant. Therefore, we can write \eqref{eqn:error} as
\begin{equation}\label{eqn:error_2}
\tilde{\mathbf{e}}_{ij}^b = \tilde{\mathbf{e}}_{ij}^{1b} + \mathbf{c}
\end{equation}
\noindent
where $\mathbf{c}$ is a constant defined as
\begin{align}
\mathbf{c} = \sum_{p=1}^{b-1} \tilde{\mathbf{e}}^{p(b-p)}_{ij} + \sum_{p=2}^{b}\tilde{\mathbf{e}}^{p(b-p+1)}_{ij}
\end{align}
Taking Fourier transform of \eqref{eqn:error_2}, we get the Fourier transformed error $\underline{\tilde{\mathbf{e}}}_{ij}^b$ as
\begin{align}\label{eqn:err}
\underline{\tilde{\mathbf{e}}}_{ij}^b &= \mathbf{F}_1\tilde{\mathbf{e}}_{ij}^{1b} + \mathbf{F}_1\mathbf{c}\nonumber\\
		  &= \mathbf{F}_1.\mathbf{A}_2(\mathbf{C}_{\tilde{x}_i^1}\hat{\mathbf{h}}_j^b - \mathbf{C}_{\tilde{x}_j^1}\hat{\mathbf{h}}_i^b) + \underline{\mathbf{c}}\nonumber\\
		  &= \mathbf{F}_1\mathbf{A}_2\mathbf{F}_2^{-1}\mathbf{F}_2(\mathbf{C}_{\tilde{x}_i^1}\hat{\mathbf{h}}_j^b - \mathbf{C}_{\tilde{x}_j^1}\hat{\mathbf{h}}_i^b) + \underline{\mathbf{c}}	  
\end{align}
\noindent
where $\mathbf{F}_1$ and $\mathbf{F}_2$ denote the DFT matrix of size $L_b\times L_b$ and $(2Lb-1)\times (2L_b-1)$, respectively. Rewriting \eqref{eqn:err}, we get
\begin{align}\label{eqn:err_hb}
 \underline{\tilde{\mathbf{e}}}_{ij}^b  &= \mathbf{F}_1\mathbf{A}_2\mathbf{F}_2^{-1}(\underline{\tilde{\mathbf{x}}}_i^{1}.*\underline{\hat{\mathbf{h}}}_j^{b} - \underline{\tilde{\mathbf{x}}}_j^{1}.*\underline{\hat{\mathbf{h}}}_i^{b}) + \underline{\mathbf{c}}\nonumber\\
					&= \mathbf{B}\underline{\mathbf{e}}_{ij}^{1b} + \underline{\mathbf{c}}
\end{align} 
where
$\mathbf{B} = \mathbf{F}_1\mathbf{A}_2\mathbf{F}_2^{-1}$, $\underline{\tilde{\mathbf{x}}}_i^1$ is the Fourier transform of the first block of the $i$-th channel TRF data $\tilde{x}_i^1$ and $\underline{\hat{\mathbf{h}}}_i^b$ denotes the Fourier transform of $\hat{\mathbf{h}}_i^b$ of length $2L_b-1$. As convolution in the time-domain becomes multiplication in the frequency-domain, we can write, for example, $\mathbf{F}_2( \mathbf{C}_{\tilde{x}_i^1}\hat{\mathbf{h}}_j^b )$ as $\underline{\tilde{\mathbf{x}}}_i^{1}.*\underline{\hat{\mathbf{h}}}_j^{b}$, where `$.*$' denotes the element-wise multiplication operation. Subsequently, underbar with any quantity will define its Fourier transform.

Now, the cost function for the $b$-th block for estimating $\underline{\hat{\mathbf{h}}}^b$ can be defined as
\begin{align}\label{eqn:cost_function}
  J^b &= \sum_{i=1}^{M-1} \sum_{j=i+1}^{M} \underline{\tilde{\mathbf{e}}}_{ij}^{bH} \underline{\tilde{\mathbf{e}}}_{ij}^b
\end{align}
Here `$H$' denotes the hermitian operation. An estimate of the $b$-th block TRF, $\underline{\hat{\mathbf{h}}}^b$ can be obtained by minimizing the cost function $J^b$ as
\begin{equation}\label{eqn:arg}
\underline{\hat{\mathbf{h}}}^b  = arg_{\underline{\mathbf{h}}^b} \min J^b, \mbox{subject~to~}
||\underline{\hat{\mathbf{h}}}|| = 1
\end{equation}
where `$||\cdot||$' denotes the $l_2$-norm and
\begin{equation}\label{eqn:total_trf}
\underline{\hat{\mathbf{h}}}(m) =
\begin{bmatrix}
\underline{\hat{\mathbf{h}}}^{1T}(m) & \underline{\hat{\mathbf{h}}}^{2T}(m)  & \cdots  & \underline{\hat{\mathbf{h}}}^{bT}(m)
\end{bmatrix}^T
\end{equation}
with 
\begin{multline}
\underline{\hat{\mathbf{h}}}^{b}(m) =
\begin{bmatrix}
\underline{\hat{\mathbf{h}}}_{1}^b(m) & \underline{\hat{\mathbf{h}}}_{2}^b(m)  & \cdots  & \underline{\hat{\mathbf{h}}}_{M}^b(m)
\end{bmatrix}
\end{multline}
In what follows, we derive a variable step-size multichannel FLMS algorithm for the solution of \eqref{eqn:arg}.

Substituting \eqref{eqn:err_hb} into \eqref{eqn:cost_function}, we get
\begin{align}\label{eqn:simple_cost}
  J^b &= \sum_{i=1}^{M-1} \sum_{j=i+1}^{M}\left(\mathbf{B}\underline{\mathbf{e}}_{ij}^{1b} + \underline{\mathbf{c}}\right)^H\left(\mathbf{B}\underline{\mathbf{e}}_{ij}^{1b} + \underline{\mathbf{c}}\right)\nonumber\\
  &= \sum_{i=1}^{M-1} \sum_{j=i+1}^{M} \left(\underline{\mathbf{e}}_{ij}^{1bH}\mathbf{B}^H\mathbf{B}\underline{\mathbf{e}}_{ij}^{1b}  + \underline{\mathbf{e}}_{ij}^{1bH}\mathbf{B}^H\underline{\mathbf{c}} + \underline{\mathbf{c}}^H\mathbf{B}\underline{\mathbf{e}}_{ij}^{1b} + \underline{\mathbf{a}}\right)
\end{align}
where $\underline{\mathbf{a}} = \underline{\mathbf{c}}^H\underline{\mathbf{c}}$ is a constant.
Taking the gradient of \eqref{eqn:simple_cost} with respect to the $b$-th block TRF of the $k$-th channel, we get
\begin{align}\label{eqn:grad_up}
 \frac{\partial J^b}{\partial \underline{\hat{\mathbf{h}}}_{k}^{b*}} &= \frac{\partial }{\partial \underline{\hat{\mathbf{h}}}_{k}^{b*}} \bigg[ \sum_{i=1}^{M-1} \sum_{j=i+1}^{M} (\underline{\mathbf{e}}_{ij}^{1bH}\mathbf{B}^H\mathbf{B}\underline{\mathbf{e}}_{ij}^{1b} + \underline{\mathbf{e}}_{ij}^{1bH}\mathbf{B}^H\underline{\mathbf{c}} + \nonumber \\ &~~~~~~~~~~~~\underline{\mathbf{c}}^H\mathbf{B}\underline{\mathbf{e}}_{ij}^{1b} + \underline{\mathbf{a}})\bigg ] \nonumber\\
                                              &= \sum_{i=1}^{k-1}\left(\underline{\tilde{\mathbf{x}}}_{i}^{1*}.*\mathbf{B}^H\mathbf{B}\underline{\mathbf{e}}_{ik}^{1b} + \underline{\tilde{\mathbf{x}}}_{i}^{1*}.*\mathbf{B}^H\underline{\mathbf{c}}\right)\nonumber\\
&~~- \sum_{j=k+1}^{M}\left(\underline{\tilde{\mathbf{x}}}_{j}^{1*}.*\mathbf{B}^H\mathbf{B}\underline{\mathbf{e}}_{kj}^{1b} + \underline{\tilde{\mathbf{x}}}_{j}^{1*}.*\mathbf{B}^H\underline{\mathbf{c}}\right)\nonumber\\
											&= \sum_{i=1}^{M}\underline{\tilde{\mathbf{x}}}_i^{1*}.*\mathbf{B}^H(\mathbf{B}\underline{\mathbf{e}}_{ik}^{1b} + \underline{\mathbf{c}})\nonumber\\
											&= \sum_{i=1}^{M}\underline{\tilde{\mathbf{x}}}_i^{1*}.*\mathbf{B}^H\underline{\tilde{\mathbf{e}}}_{ik}^b,~~~~ k = 1,2,\cdots ,M
\end{align}
Here `*' denotes the conjugate operation. The update equation of the \textit{b}MCFLMS algorithm for the $b$-th block of the RF data is given by 
\begin{equation}\label{eqn:update_eqn}
\underline{\hat{\mathbf{h}}}^b(m+1) = \underline{\hat{\mathbf{h}}}^b(m) - \mu^b(m) \nabla_b J^b(m)|_{\underline{\mathbf{h}} = \hat{\underline{\mathbf{h}}}(m)}, b = 1, 2,\cdots, B
\end{equation}
where,
\begin{align}
\nabla_b J^b(m) &= \frac{\partial J^b(m)}{\partial \underline{\hat{\mathbf{h}}}^{b*}(m)} \nonumber\\
&=
\begin{bmatrix}
\frac{\partial J^b(m)}{\partial \underline{\hat{\mathbf{h}}}_{1}^{b*}(m)} & \frac{\partial J^b(m)}{\partial \underline{\hat{\mathbf{h}}}_{2}^{b*}(m)}  & \cdots  & \frac{\partial J^b(m)}{\partial \underline{\hat{\mathbf{h}}}_{M}^{b*}(m)}
\end{bmatrix}
\end{align}
Here, $\underline{\hat{\mathbf{h}}}^b(m)$ denotes the $m$-th iteration estimate of $\mathbf{h}^b$ and $\mu^b(m)$ is the variable step-size (VSS) for the $b$-th block. The step-size is adapted so that the distance between $\underline{\hat{\mathbf{h}}}^b(m+1)$ and $\underline{\hat{\mathbf{h}}}^b(m)$ becomes minimum at each iteration and for noise-free case it is given by (see \cite{gaubitch2006generalized} for details)
\begin{equation}
\mu^b(m) = \frac{\hat{\underline{\mathbf{h}}}^{bT}(m)}{||\nabla_b J^b(m)||^2}\nabla_b J^b(m) - \frac{\underline{\bf{h}}^{bT}(m)}{||\nabla_b J^b(m)||^2}\nabla_b J^b(m)
\end{equation}
Here `T' denotes the transpose operation. The problem with this equation is that we need to know the true TRF of the $b$-th block to calculate $\mu^b(m)$ which is not known beforehand. Unlike in \cite{hasan2006damped},  $\underline{\mathbf{h}}^{b}$ and $\nabla_b J^b(m)$ are not orthogonal because $\nabla_b J^b(m)$ is not only a function of $\underline{\mathbf{h}}^{b}$, but also of other blocks up to the $b$-th block. Therefore, we use the TRFs and the gradient of the cost function up to the $b$-th block to calculate the step-size as
\begin{equation}\label{eqn:mu}
\mu^b(m) = \frac{\hat{\underline{\mathbf{h}}^{T}}(m)}{||\nabla J^b(m)||^2}\nabla J^b(m) - \frac{\underline{\mathbf{h}}^{T}(m)}{||\nabla J^b(m)||^2}\nabla J^b(m)
\end{equation}
where,
\begin{multline}\label{eqn:gradient}
\nabla J^b(m) = \frac{\partial J^b(m)}{\partial \underline{\hat{\mathbf{h}}}^*(m)}\\
=
\begin{bmatrix}
(\frac{\partial J^b(m)}{\partial \underline{\hat{\mathbf{h}}}^{1*}(m)})^T & (\frac{\partial J^b(m)}{\partial \underline{\hat{\mathbf{h}}}^{2*}(m)})^T  & \cdots  & (\frac{\partial J^b(m)}{\partial \underline{\hat{\mathbf{h}}}^{b*}(m)})^T
\end{bmatrix}^T
\end{multline}
Now, the second term of \eqref{eqn:mu} becomes zero as the true TRFs vector, \underline{\bf{h}} formed by concatenating the true TRFs of all the blocks up to the $b$-th block is orthogonal to $\nabla J^b(m)$. The step-size in \eqref{eqn:mu} then becomes
\begin{equation}\label{eqn:mu_1}
\mu^b(m) = \frac{\hat{\underline{\mathbf{h}}}^{T}(m)}{||\nabla J^b(m)||^2}\nabla J^b(m)
\end{equation}
To evaluate \eqref{eqn:gradient} and \eqref{eqn:mu_1}, we also need to derive the gradients of \eqref{eqn:cost_function} with respect to other blocks $q$, where $q \neq b$. To this end, as we are taking gradient with respect to $\hat{\underline{h}}^q$, we can consider other parts of \eqref{eqn:error} which do not depend on $\hat{\underline{h}}^q$ as constant. From \eqref{eqn:error}, we can write
\begin{align}\label{eqn:error2}
\tilde{\mathbf{e}}^b_{ij} &= \tilde{\mathbf{e}}^{(b-q)q} + \tilde{\mathbf{e}}^{(b-q+1)q} + \sum_{p=1,p \neq b-q}^{b-1}\tilde{\mathbf{e}}^{p(b-p)} \nonumber\\
		 &~~+ \sum_{p=1,p \neq b-q+1}^{b}\tilde{\mathbf{e}}^{p(b-p+1)}\nonumber\\
		 &= \tilde{\mathbf{e}}^{(b-q)q} + \tilde{\mathbf{e}}^{(b-q+1)q} + \mathbf{c}_1
\end{align}
where the constant $\mathbf{c}_1$ is defined as 
\begin{align}
\mathbf{c}_1 = \sum_{p=1,p \neq b-q}^{b-1}\tilde{\mathbf{e}}^{p(b-p)} + \sum_{p=1,p \neq b-q+1}^{b}\tilde{\mathbf{e}}^{p(b-p+1)}
\end{align}
Taking the Fourier transform of \eqref{eqn:error2}, we get
\begin{equation}
\underline{\tilde{\mathbf{e}}}^b_{ij} = \mathbf{B}_1\underline{\mathbf{e}}^{(b-q)q} + \mathbf{B}\underline{\mathbf{e}}^{(b-q+1)q} + \underline{\mathbf{c}}_1
\end{equation}
where
$\mathbf{B}_1 = \mathbf{F}_1\mathbf{A}_1\mathbf{F}_2^{-1}$. \\
Now the cost function in \eqref{eqn:cost_function} becomes 
\begin{align*}
  J^b &= \sum_{i=1}^{M-1} \sum_{j=i+1}^{M} \underline{\tilde{\mathbf{e}}}_{ij}^{bH} \underline{\tilde{\mathbf{e}}}_{ij}^b\\
  &= \sum_{i=1}^{M-1} \sum_{j=i+1}^{M}\big(\mathbf{B}_1\underline{\mathbf{e}}_{ij}^{(b-q)q} + \mathbf{B}\underline{\mathbf{e}}_{ij}^{(b-q+1)q} + \underline{\mathbf{c}}_1)^H\\ 
   &~~~~~~~~~~~~~~~\big(\mathbf{B}_1\underline{\mathbf{e}}_{ij}^{(b-q)q} + \mathbf{B}\underline{\mathbf{e}}_{ij}^{(b-q+1)q} + \underline{\mathbf{c}}_1)
\end{align*}
Taking gradient with respect to the conjugate of $\underline{\hat{\mathbf{h}}}_{k}^q$, where $q \neq b$, we get
\begin{equation}\label{eqn:grad_q}
 \frac{\partial J^b}{\partial \underline{\hat{\mathbf{h}}}_{k}^{q*}} 
                        = \sum_{i=1}^{M}(\underline{\tilde{\mathbf{x}}}_{i}^{(b-q)*} .*\mathbf{B}_1^H \underline{\tilde{\mathbf{e}}}_{ik}^b + \underline{\tilde{\mathbf{x}}}_{i}^{(b-q+1)*}.* \mathbf{B}^H\underline{\tilde{\mathbf{e}}}_{ik}^b)\\
\end{equation}
Using \eqref{eqn:grad_q} with \eqref{eqn:gradient}, the VSS $\mu^b(m)$ in \eqref{eqn:mu_1} can now be calculated and the TRFs are updated using \eqref{eqn:update_eqn}. The TRF up to the $b$-th block is normalized to avoid the trivial zero solution after each update, i.e.,
\begin{equation}\label{eqn:normalize}
\underline{\hat{\mathbf{h}}}(m+1) = \frac{\underline{\hat{\mathbf{h}}}(m+1)}{||\underline{\hat{\mathbf{h}}}(m+1)||}
\end{equation}
Here unity norm constraint is applied on all the blocks of TRFs up to the $b$-th block among the total $B$ blocks of TRFs, because the cost function $J^b$ contains all the TRFs up to $\underline{\hat{\mathbf{h}}}^b$. Finally, the inverse Fourier transform of the estimate obtained using \eqref{eqn:update_eqn} will result in an estimate of the $b$-th block of the TRF. Executing \eqref{eqn:update_eqn} for all the blocks $b = 1, 2, \cdots, B,$ an estimate of the full TRF can be obtained from \eqref{eqn:total_trf}.
%\subsection{\bf{Lateral Blocking Of TRF Channels}}
%To facilitate faster convergence we propose the following type of lateral blocking of the TRF channels. First we randomized the order of M number of channels. Then we divided the randomized channels into K number of slots each of P number of randomized channel, where, \\\\
%$ K = \dfrac{M}{P}$\\
%Then we used the MCFLMS algorithm on each slot upto a particular number of iterations. We repeated the whole process for a particular number of times.

%%%%%%%%%%%%%%%%%%%%%%%%%%%%%%%%%%%%%%%%%%%%%%%%%%%%%%%%%%%%%%%%%%%%%%%%%%%%%%%%%%%%%%%%%%%%%%%
\subsection{Effect of Noise on the Convergence of the Algorithm}
So far we have assumed a noise-free case for the adaptive algorithm. However, the presence of noise may not be avoided in practice. It is well-known that noise has a significant impact on the convergence of the crossrelation based adaptive algorithms \cite{haque2008noise}, \cite{liao2015analysis}, \cite{haque2007energy}, \cite{hasan2005improving}, \cite{hasan2006analyzing}. To stop misconvergence of this type of algorithms due to the effect of noise, a spectral constraint  is proposed in \cite{haque2008noise} where it is assumed that the acoustic channel impulse response is spectrally flat. This assumption, therefore, is not valid for TRFs. The time-domain approach described in \cite{hasan2016blind} uses damped variable step-size as described in \cite{hasan2006damped}, gradient averaging, and \textit{$l_1$}-norm constraint to prevent misconvergence. The \textit{$l_1$}-norm constraint gives some sort of robustness against noise when the RF data is sparse. But unfortunately sparsity of \emph{in-vivo} data cannot be guaranteed.  In \cite{liao2015analysis} a modified cost function has been proposed where it is assumed that the additive noise in different channels have the same variance, a condition that may not be satisfied in practice. Thus, none of these techniques are generalized to address the problem of misconvergence. In this work, we develop a generalized approach to get rid of this problem and thereby make the adaptive algorithm robust to noise.

As described in the previous section, we attempt to make the error function described in \eqref{eqn:full_err} zero block by block in the proposed algorithm. In noisy case, \eqref{eqn:full_err} becomes 
\begin{align}
e_{ij}(n) &= [x_i(n) + n_i(n)]*h_j(n) - [x_j(n) + n_j(n)]*h_i(n) \nonumber\\
				&= [x_i(n)*h_j(n) - x_j(n)*h_i(n)] + \nonumber \\
				&~~~[n_i(n)*h_j(n) - n_j(n)*h_i(n)]  \nonumber\\
			    &= e^s_{ij}(n) + e^n_{ij}(n)
\end{align}
where $n_i(n)$ denotes the additive noise in the $i$-th channel, $e^s_{ij}(n)$ is the error due to noiseless data and $e^n_{ij}(n)$ is the error due to noise. Therefore, in noisy case the gradient in \eqref{eqn:grad_up} will also have two components -- one due to $e^s_{ij}(n)$ and the other is for $e^n_{ij}(n)$. However, as the noise power is generally lower than the desired signal power, in the beginning, the signal gradient will be higher than the noise gradient. Thus $e^s_{ij}(n)$ will reduce to a lower value faster than $e^n_{ij}(n)$. And at a certain instant of iteration, the two error values will be comparable. After this point, the noise gradient dominates and causes the solution to misconverge. In order to prevent the algorithm from misconverging, we have to introduce a constraint in the estimation process that somehow restrains the noisy gradient.

\begin{figure*}[ht!]
\centering
\includegraphics[width = 7.2 in, height = 2.5 in]{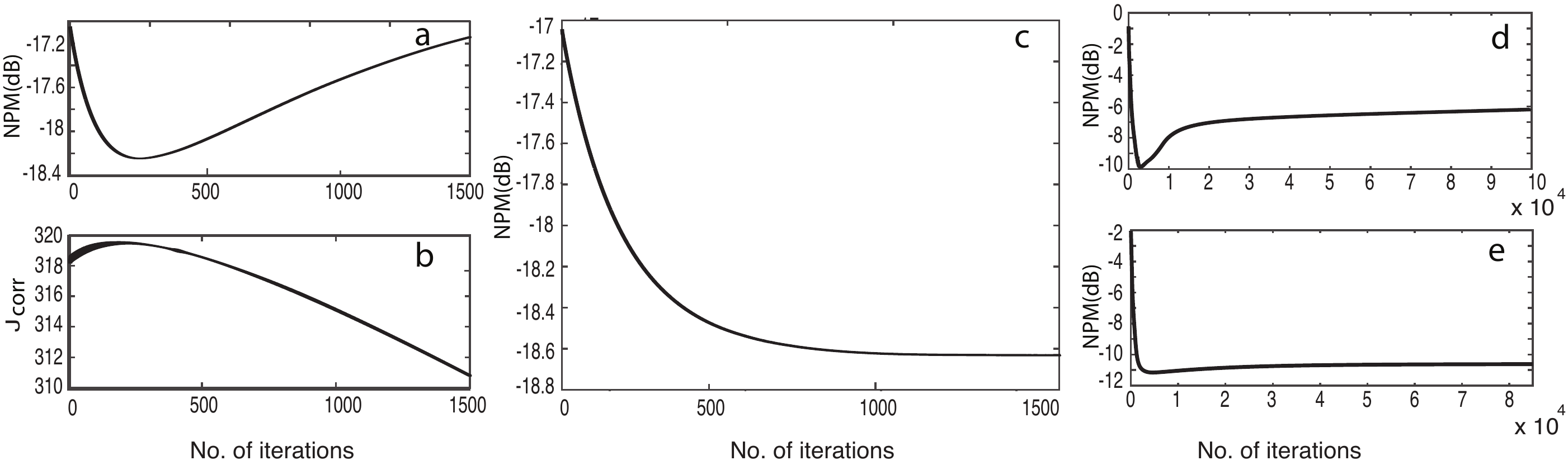}
 \caption{Effect of noise on the convergence of the \textit{b}MCFLMS algorithm (a) Behavior of the NPM curve around the misconvergence point for the first block with SNR $= 30$ dB. (b) Behavior of the correlation cost function around the misconvergence point for the first block. (c) Behavior of the NPM curve around the misconvergence point for the first block with correlation constraint ($\xi = 3$e-$7$, $\rho = 2.55$, $\gamma = 2.3$). (d) Misconvergence phenomenon of the second block of TRFs for $B = 2$ for with no additive white noise in the data.  (e) Misconvergence Problem Solved for the Second Block of TRFs for $B = 2$ using correlation constraint.}
 \label{noise_performance}
\end{figure*}

Consider the following correlation between the RF data $x_i(n)$ and estimated TRF $\hat{h}_i(n)$:
\begin{equation}\label{eqn:correlation2}
r_i'(n) = x_i(n)*\hat{h}_i(-n)
\end{equation}
Using \eqref{eqn:model}, \eqref{eqn:correlation} can be expressed as
\begin{align}\label{eqn:matched_filter}
r_i'(n) &= [s(n)*h_i(n) + v_i(n)]*\hat{h}_i(-n) \nonumber\\
       &= s(n)*h_i(n)*\hat{h}_i(-n) + v_i(n)*\hat{h}_i(-n) \nonumber\\
	   &= s(n)*{h_i(n)*\hat{h}_i(-n)} \nonumber\\
	   &= s(n)*r_{h_i}(n)
\end{align}
where
the correlation of noise with TRF, i.e., $v_i(n)*\hat{h}_i(-n)$ is assumed zero and
\begin{align}
r_{h_i}(n) = h_i(n)*\hat{h}_i(-n)
\end{align}
From filtering point of view, \eqref{eqn:matched_filter} can be viewed as if $\hat{h}_i(-n)$ is filtered by $h_i(n)$ and the filter output is further filtered by $s(n)$. Here only $\hat{h}_i(n)$ is changing with iteration.  Up to the misconvergence point $\hat{h}_i(n)$ is getting closer in shape to the true TRF $h_i(n)$ and after the misconvergence point, it deviates from the shape of $h_i(n)$. From the concept of matched filter \cite{turin1960introduction}, we know that a filter passes maximum energy at its output if the input signal is of the same shape of the filter impulse response. Therefore, the energy in $r_{h_i}(n)$ increases as $\hat{h}_i(n)$ gets closer to the misconvergence point, but it decreases after the misconvergence point. As a result, the energy in $r_i'(n)$ reaches its maximum value at the misconvergence point. In this paper, we exploit this phenomenon to prevent misconvergence. With a little modification of \eqref{eqn:correlation}, consider the following convolution:
\begin{equation}\label{eqn:convln}
r_i(n) = x_i(n)*\hat{h}_i(n)
\end{equation}
It is obvious that $r_i'(n)$ and $r_i(n)$ both have the same power spectrum but different phase spectrum. Therefore, we can use \eqref{eqn:convln} instead of \eqref{eqn:correlation2} while using its energy as constraint to prevent misconvergence. Since we estimate TRFs block by block, we cannot use the full length convolution as described in \eqref{eqn:convln}, rather we will follow the block-based convolution as described in the previous section. In a similar manner to \eqref{eqn:blocked_convolution}, for the $b$-th block of the total convolution length defined in \eqref{eqn:convln}, we can write
\begin{equation}\label{eqn:correlation}
\mathbf{r}_i^b = \sum_{p=1}^{b-1}\mathbf{A}_1\mathbf{C}_{x_i^p}\hat{\mathbf{h}}_i^{(b-p)} + \sum_{p=1}^{b}\mathbf{A}_2\mathbf{C}_{x_i^p}\hat{\mathbf{h}}_i^{(b-p+1)} 
\end{equation}
To use the energy in $\mathbf{r}_i^b$ as constraint, consider the following cost function for the $b$-th block of data for $i = 1, 2, \cdots, M$:
\begin{equation} \label{eqn:corr_cons}
J_{corr}^b = \sum_{i=1}^{M}\underline{\mathbf{r}}_i^{bH}~\underline{\mathbf{r}}_i^{b}
\end{equation}
In order to show that $J_{corr}^b$ becomes maximum at the misconvergence point graphically, the simulation phantom data as described later in the result section was used. In Fig. \ref{noise_performance}(a) the misconvergence phenomenon is shown on the simulation data with  $30$ dB SNR. As shown in Fig. \ref{noise_performance}(b), the first block correlation cost function  $J_{corr}^1$ is maximum at the misconvergence point and then it decreases. Therefore, the misconvergence due to noise can be avoided if we minimize the $b$-th block cost function $J^b$ in \eqref{eqn:cost_function} while at the same time maximize $J_{corr}^b$ or equivalently minimize $-J_{corr}^b$. Adding \eqref{eqn:corr_cons} as constraint to our previous cost function in \eqref{eqn:cost_function} gives
\begin{equation}\label{eqn:constraint}
J^{bt}(m) = J^b(m) - \psi(m) J_{corr}^b(m)
\end{equation}
where $\psi(m)$ is the Lagrange multiplier, also known as the coupling factor. In general, for any block $b$, the gradient of the cost function with respect to  $\underline{\mathbf{h}}_k^{b*}$ can be written as
\begin{equation}\label{eqn:J^bt}
\frac{\partial J^{bt}}{\partial \underline{\mathbf{\hat{h}}}_{k}^{b*}} = \sum_{i=1}^{M}\underline{\mathbf{x}}_i^{1*}.*\mathbf{B}^H\underline{\mathbf{e}}_{ik}^b - \psi(m) \underline{\mathbf{x}}_k^{1*}.*\mathbf{B}^H\underline{\mathbf{r}}_k^{b}
\end{equation}
Considering the correlation constraint with \eqref{eqn:grad_q}, we get
\begin{multline}\label{eqn:J^bqt}
 \frac{\partial J^{bt}}{\partial \underline{\mathbf{\hat{h}}}_{k}^{q*}} 
                        = \sum_{i=1}^{M}(\underline{\mathbf{x}}_{i}^{(b-q)*}.*\mathbf{B}_1^H\underline{\mathbf{e}}_{ik}^b + \underline{\mathbf{x}}_{i}^{(b-q+1)*}.*\mathbf{B}^H\underline{\mathbf{e}}_{ik}^b)\\
- \psi(m) (\underline{\mathbf{x}}_{k}^{(b-q)*}.*\mathbf{B}_1^H\underline{\mathbf{r}}_k^{b} + \underline{\mathbf{x}}_{k}^{(b-q+1)*}.*\mathbf{B}^H\underline{\mathbf{r}}_k^{b})
\end{multline}
Here the update process of the TRFs is the same as described in the previous section with a change in the gradient. Now using the gradient from \eqref{eqn:J^bt} and \eqref{eqn:J^bqt} in \eqref{eqn:gradient}, \eqref{eqn:total_trf} and \eqref{eqn:mu_1}, we can calculate $\mu^b(m)$. Next, we update the $b$-th block TRFs using \eqref{eqn:update_eqn}.

The coupling factor $\psi(m)$ in \eqref{eqn:constraint} should be so chosen that it gives a smaller value for a higher value of $J^b(m)$ and a larger value for a lower value of $J^b(m)$. This is due to the fact that initially the value of $J^b(m)$ will be high and the gradient due to noiseless data is dominant. Therefore, initially the coupling factor should be of a small value to facilitate unconstrained update of the TRF due to the dominant gradient for the noiseless data. The update equation described in \eqref{eqn:update_eqn} makes the cost function decrease even after the misconvergence point and at this point the gradient due to noise becomes comparable to signal gradient. For this reason,  $\psi(m)$ should be so chosen that it increases with the decrease of $J^b(m)$  for a small value of $J^b(m)$. As it crosses the misconvergence point, the higher value of $\psi(m)$ makes the noise effect compensation stronger.  A suitable expression for the coupling factor is empirically obtained as 
\begin{equation}
\psi(m) = \xi(|\rho~log_{10}(J^b(m))|)^\gamma
\end{equation}
Here $\gamma$ determines the sensitivity of $\psi(m)$ to decrease in the value of $J^b(m)$. A higher value of $\gamma$ means that $\psi(m)$ will increase highly for a small decrease of $J^b(m)$.  Fig. \ref{noise_performance}(c) shows that the misconvergence problem is solved after adding the constraint described in \eqref{eqn:constraint}. Misconvergence phenomenon can also emerge from  estimation noise. For example, as we are using the estimated TRFs of the first block instead of true TRFs  while estimating the TRFs of the second block, actually we are adding estimation noise to the process. This phenomenon is shown in Fig. \ref{noise_performance}(d). The misconvergence becomes stronger as the number of blocks increases. Fig. \ref{noise_performance}(e) shows that our proposed constraint also works against the estimation noise. Figs. \ref{noise_performance}(c) and (e) also show that the proposed constraint helps to achieve better NPM than that of the misconverging point.

The proposed \textit{b}MCFLMS algorithm is summarized in Table~I.

\begin{table}[h!]  \label{tab: b-mcflms}
  %  \centering
\caption{Constrained \textit{b}MCFLMS Algorithm}
  \rule{8.5cm}{0.4pt}
  \begin{enumerate}
  \item[\bf{Step 1}]
    \begin{enumerate}
    \item[.] Set total block number $B$ and appropriate value for $\psi$, $\rho$ and $\gamma$
    \item[.] Initialize the $i$-th channel TRF, $\mathbf{\hat{h}}_i = [1~0_{1 \times (L-1)}]^T$ for $i = 1, 2,\cdots, M$
    \end{enumerate}
    
\item[\bf{Step 2}]
    \begin{enumerate}
    \item[.] Set current block number $b = 1$
    \end{enumerate}
    
\item[\bf{Step 3}]
    \begin{enumerate}
    \item[.] Set iteration index $m = 1$
    \end{enumerate}
    
 \item[\bf{Step 4}]
    \begin{enumerate}
    \item[.] Calculate the error functions for $b$-th block using \eqref{eqn:last_err}, \eqref{eqn:first_err}, \eqref{eqn:error} and \eqref{eqn:err_hb}
    \item[.] Calculate correlation of estimated TRF with RF data using \eqref{eqn:correlation}
    \end{enumerate}
    
\item[\bf{Step 5}]
    \begin{enumerate}
    \item[.] Calculate $\frac{\partial J^{bt}}{\partial \underline{\mathbf{\hat{h}}}_{k}^{b*}}$ according to \eqref{eqn:J^bt}
    \item[.] Calculate $\frac{\partial J^{bt}}{\partial \underline{\mathbf{\hat{h}}}_{k}^{q*}}$ according to \eqref{eqn:J^bqt}
    \item[.] Calculate step-size for $b$-th block and $m$-th iteration $\mu^b(m)$ using $J^{bt}$ instead of $J^b$ in  \eqref{eqn:gradient}, \eqref{eqn:total_trf} and \eqref{eqn:mu_1}
    \end{enumerate}  
       
\item[\bf{Step 6}]
    \begin{enumerate}
    \item[.] Update $\underline{\mathbf{\hat{h}}}$ using \eqref{eqn:update_eqn}
    \item[.] Normalize $\underline{\mathbf{\hat{h}}}$ according to \eqref{eqn:normalize}
    \end{enumerate}
    
\item[\bf{Step 7}]
    \begin{enumerate}
    \item[.] If $m$ is less than required iterations, set $m = m + 1$ and go to step $4$
    \item[.] Else set $b = b + 1$, $m = 1$ and go to step $3$
    \end{enumerate}    
  \end{enumerate}
  \rule{8.5cm}{0.4pt}
\end{table}

%%%%%%%%%%%%%%%%%%%%%%%%%%%%%%%%%%%%%%%%%%%%%%%%%%%%%%%%%%%%%%%%%%%%%%%%%%%%%%%
\section{Resolution Improvement Using Estimated Missing Data}
It is described in section III that incomplete data acquisition in ultrasound imaging results in $B$ blocks of usable errors as given by \eqref{eqn:error}. While estimating a TRF $h_i(n)$ of a particular channel $i$, we have $L$ unknown coefficients and thus we have $L\times M$ number of unknowns in total for $M$ channels. The use of error blocks up to the $B$-th block gives us $L\times M(M-1)/2$ equations to estimate these $L\times M$ coefficients. However, if we can use an extra error block, i.e., the $(B+1)$-th error block, we will have additional $L_b\times M(M-1)/2$ equations to estimate the TRFs. Therefore, incorporation of these extra equations into the estimation algorithm is expected to provide more robustness against noise and may also help improve the resolution of the ultrasound image further. 

Now consider the following convolution matrix of the $i$-th channel TRF data: 
\begin{multline*}
\mathbf{C}_{x_{i}} = \\
\left[\begin{smallmatrix}
   {x_{i}(n)} & 0  & \cdots & 0  \\
   {x_{i}(n-1)} & {x_{i}(n)}  & \cdots & 0  \\
 \vdots  &  \vdots  &  \ddots &  \vdots \\
   {x_{i}(n-L+1)} & {x_{i}(n-L+2)}   &  \cdots & x_{i}(n) \\ \hline
   {x_{i}(n-L)} & {x_{i}(n-L+1)}  &  \cdots  & x_{i}(n-1)  \\
 \vdots  &  \vdots   &  \ddots &  \vdots  \\
   {x_{i}(n-L-L_s+2)} & {x_{i}(n-L-L_s+3)} & \cdots  & {x_{i}(n-L_s+1)} \\
   0 & {x_{i}(n-L-L_s+2)} & \cdots & {x_{i}(n-L_s)} \\
 \vdots  &  \vdots   &  \ddots &  \vdots \\
   0 & 0 & \cdots  & {x_{i}(n-L_b+1)} \\ \hline
   0 & 0 &  \cdots & {x_{i}(n-L_s)} \\
   \vdots  &  \vdots   &  \ddots &  \vdots \\
   0 & 0 & \cdots & {x_{i}(n-L-L_s+2)} \\
\end{smallmatrix}\right]
\end{multline*}
So far we have been using the convolution matrix up to the first drawn line, i.e., $L$ number of rows. Now in order to introduce redundancy in the estimation equations, we wish to use the next block of the convolution matrix up to the second drawn line that will give us the $(B+1)$-th block of the error function. However, due to the nature of ultrasound data acquisition, the extra block of data needed for the $(B+1)$-th block of $\mathbf{C}_{x_{i}}$ is not available from the transducers. To estimate this block of missing data, we can use the estimated TRFs from the \textit{b}MCFLMS algorithm and the PSF can be estimated using the R-MINT algorithm as shown in \cite{hasan2016blind}. Again if we observe the lower part of the convolution matrix $\mathbf{C}_{x_{i}}$, i.e., the part below the second drawn line, we find that this block contains mostly estimated data. Therefore, if we use the $(B+2)$-th block of the error function, it may deteriorate the estimation accuracy. Hence, the $(B+2)$-th block of the error function is not used in the proposed \textit{md-b}MCFLMS algorithm.
%%%%%%%%%%%%%%%%%%%%%%%%%%%%%%%%%%%%%%%%%%%%%%%%%%%%%%%%%%%%%%%%%%%%%%%%%%%%%%%
\subsection{Ultrasound Pulse or PSF Estimation}
\begin{figure}[h]
\includegraphics[scale=.45]{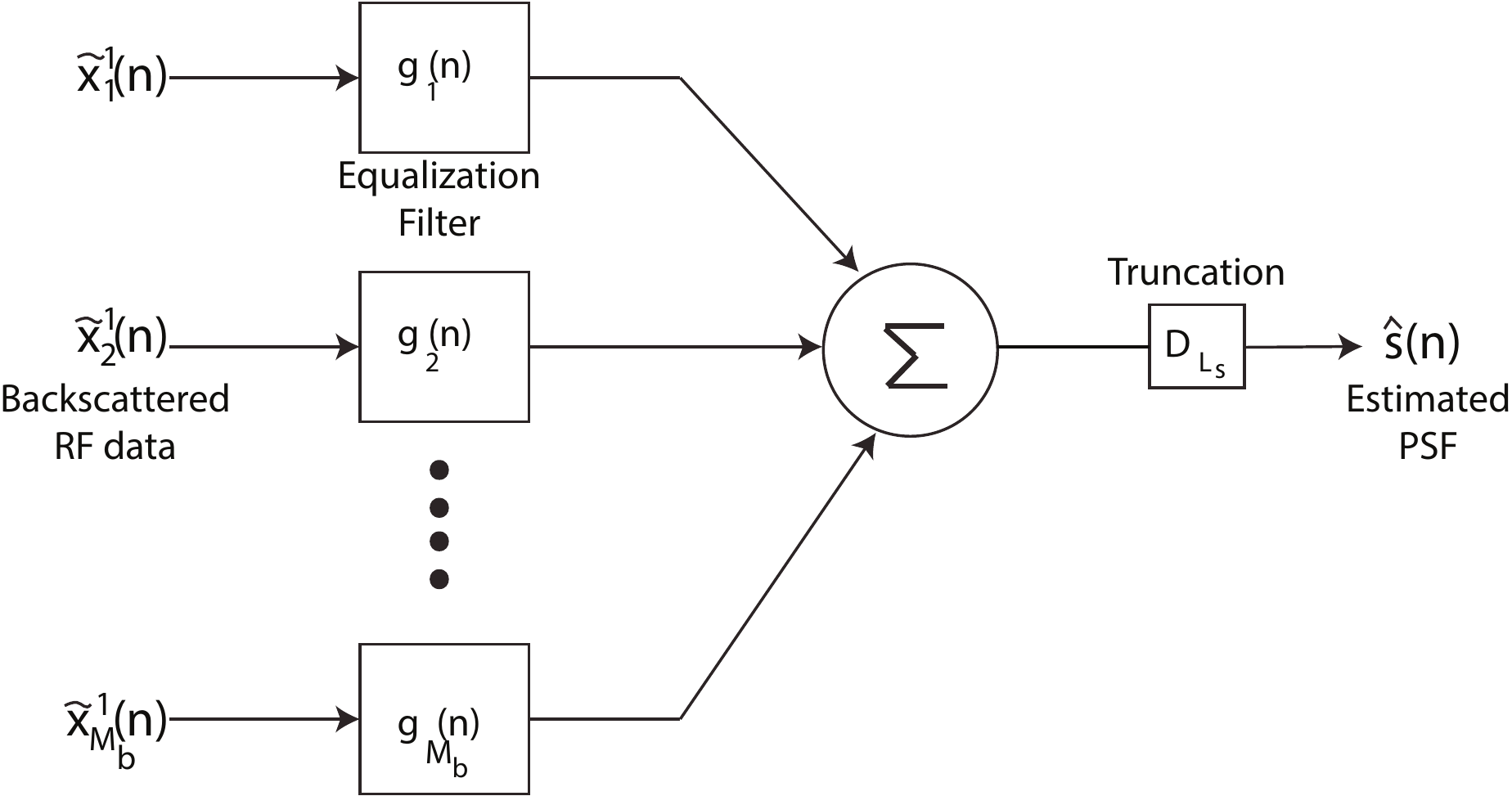}
\caption{The block diagram for the R-MINT estimation of PSF for a single block of data.}
\label{fig:MINT}
\end{figure}
In our work, we use the robust multiple-input/output inverse theorem (R-MINT) to estimate the PSF from the first block ($b = 1$) of the axially blocked RF data. Figure~\ref{fig:MINT} shows the block diagram of the algorithm. Here, the equalization filter $\bf{g}_i$ of the $i$-th channel removes the TRF effect from the RF data and gives the PSF. The PSFs from different channels are then added to reduce the estimation error. Mathematically, this can be expressed as 
\begin{equation}\label{eqn:psf}
\hat{\mathbf{s}} = \mathbf{D}_{L_s}\mathbf{C}_{\tilde{x}^1}\mathbf{g} 
\end{equation}

where
\begin{align*}
\hat{\mathbf{s}} &= 
\begin{bmatrix}
\hat{s}(n) &\hat{s}(n-1) \cdots & \hat{s}(n-L_s+1)
\end{bmatrix}^T
\\
\mathbf{D}_{L_s} &= 
\begin{bmatrix}
\bf{I}_{L_s}   & \bf{0}_{L_s\times(L_b-L_s)}
\end{bmatrix}
\\
\mathbf{C}_{\tilde{x}^1} &= 
\begin{bmatrix}
\mathbf{C}_{\tilde{x}_1^1} & \mathbf{C}_{\tilde{x}_2^1} \cdots & \mathbf{C}_{\tilde{x}_M^1}
\end{bmatrix}
,\\
\mathbf{g} &= 
\begin{bmatrix}
\mathbf{g}_1^T &\mathbf{g}_2^T \cdots  &\mathbf{g}_M^T
\end{bmatrix}^T
\end{align*}

If we can estimate $\mathbf{g}$, we can determine the PSF. Now, as shown in \eqref{eqn:model}, for noise-free case $C_{\tilde{x}^1}$ can be written as
\begin{equation}\label{eqn:conv}
\mathbf{C}_{\tilde{x}^1} = \mathbf{S}_{L0}\mathbf{H}
\end{equation}
where
\begin{align*}
\mathbf{S}_{L0} &= 
\begin{bmatrix}
\mathbf{S}_{L_b} & \mathbf{0}_{L_b\times(L_b-1)}
\end{bmatrix}\nonumber
\end{align*}

\begin{align*}
\mathbf{S}_{L_b} &= 
\begin{bmatrix}
s(n) &0 \cdots &0 \cdots &0 &\cdots &0\\
s(n-1) &s(n) &\cdots &0 &\cdots &0\\
\vdots &\vdots &\ddots &\vdots &\ddots &\vdots\\
0 &\cdots &0 &s(n-L_s+1) &\cdots &s(n)
\end{bmatrix}\nonumber
\end{align*}
and
\begin{align*}
\mathbf{H} &= 
\begin{bmatrix}
\mathbf{H}_1 & \mathbf{H}_2 \cdots & \mathbf{H}_M
\end{bmatrix}\nonumber
\end{align*}
Here $\mathbf{H}_i$ is the convolution matrix of $i$-th channel TRF. Using \eqref{eqn:psf} and \eqref{eqn:conv}, we can write
\begin{equation}\label{eqn:s_hat}
\hat{\mathbf{s}} = \mathbf{D}_{L_s}\mathbf{S}_{L0}\mathbf{Hg} 
			 = \mathbf{D}_{L_s}\mathbf{S}_{L0}\mathbf{d} 
\end{equation}
Equation \eqref{eqn:s_hat} reveals that if $\bf{d}$ is an impulse function we can extract PSF from $\mathbf{S}_{L0}$. Therefore, our estimated $\mathbf{g}$ should be such that $\bf{Hg}$ is close to an impulse function. As the equalization filter $\mathbf{g}$ removes the TRF effect from the RF data, the energy in $\mathbf{g}$ will become unbounded where TRF has lower magnitude. A regularization constraint is needed to keep the energy of the filter bounded. Now, the cost function to evaluate $\mathbf{g}$ as shown in \cite{hasan2016blind} is given by
\begin{equation}\label{eqn:equilizer}
\operatornamewithlimits{argmin}_{\mathbf{g}} ||\hat{\mathbf{H}}\mathbf{g} - \mathbf{d}||^2 + \delta||\mathbf{g}||^2
\end{equation}
Here $\delta$ is a regularization parameter and $\hat{\mathbf{H}}$ is the estimated multichannel TRF matrix. Solving \eqref{eqn:equilizer}, we get
\begin{equation}\label{eqn:g}
\mathbf{g} = (\hat{\mathbf{H}}^T\hat{\mathbf{H}} + \delta\mathbf{I})^{-1}\hat{\mathbf{H}}^T\mathbf{d}
\end{equation}
Estimation of the PSF using all the $M$ channel TRFs according to \eqref{eqn:g} is desirable. However, there may be a memory limitation involved in the direct implementation of \eqref{eqn:g}. To reduce the size of the $\hat{\mathbf{H}}$ matrix, we divide the TRFs into lateral blocks with equal number of TRFs and then estimate the PSF $\hat{\mathbf{s}}_r$ for these blocks as
\begin{equation}\label{eqn:g_r}
\mathbf{g}_r = (\hat{\mathbf{H}}^T_r\hat{\mathbf{H}}_r + \delta\mathbf{I})^{-1}\hat{\mathbf{H}}^T_r\mathbf{d}, ~~~~~~ r=1,2,\cdots,N
\end{equation}
where $N = floor(M/M_b)$ and $M_b$ is the number of TRFs in each block. Now, the PSF can be determined according to \eqref{eqn:psf} as
\begin{equation}\label{eqn:s_r}
\hat{\mathbf{s}}_r = \mathbf{D}_{L_s}\mathbf{C}_{\tilde{x}^1}^r\mathbf{g}_r
\end{equation}
Here $\mathbf{C}_{\tilde{x}^1}^r$ is the convolution matrix of the first block of RF data containing $r$ TRFs. The final estimate of the PSF is the average of all the estimated TRFs given by
\begin{equation}\label{eqn:hat_h}
\hat{\mathbf{s}} = \frac{1}{N}\sum_{r=1}^N \hat{\mathbf{s}}_r
\end{equation}

%%%%%%%%%%%%%%%%%%%%%%%%%%%%%%%%%%%%%%%%%%%%%%%%%%%%%%%%%%%%%%%%%%%%%%%%%%%%%%%%%%%%%%%%%%%
\subsection{\textit{md-b}MCFLMS Algorithm}
In this section, we will use the estimated $(B+1)$-th block of the error function to improve the resolution of the ultrasound images further. Now, using the estimated PSF and TRFs, we can estimate the $i$-th channel data as
\begin{equation}\label{eqn:data_est}
\hat{\mathbf{x}}_i = \hat{\mathbf{S}}\hat{\mathbf{h}}_i
\end{equation}
where

%\begin{align*}
%\bf{\hat{S}} &= 
%\begin{bmatrix}
%\mathbf{0}_{L_b\times(L-L_s) \mathbf{S}_{L_b}
%\end{bmatrix},
%\end{align*}

\begin{align*}
\mathbf{\hat{S}} &= 
\left[\begin{smallmatrix}
\hat{s}(n) &\cdots &0 &0 &\cdots &0\\
\hat{s}(n-1) &\hat{s}(n) &\cdots &0 &\cdots &0\\
\vdots &\vdots &\ddots &\vdots &\ddots &\vdots\\
0 &0 &\cdots &\hat{s}(n-L_s+2) &\cdots &0\\ \hline
0 &0 &\cdots &\hat{s}(n-L_s+1) &\cdots &0\\
\vdots &\vdots &\ddots &\vdots &\ddots &\vdots\\ \hline
0 &\cdots &0 &\hat{s}(n-L_s+1) &\cdots &\hat{s}(n)\\
\vdots &\vdots &\ddots &\vdots &\ddots &\vdots\\
0 &0 &\cdots &0 &\cdots &\hat{s}(n-L_s+1)
\end{smallmatrix}\right],
\end{align*}

\begin{align*}
\hat{\mathbf{h}}_i &= 
\begin{bmatrix}
\hat{h}_i(n) &\hat{h}_i(n-1) \cdots  & \hat{h}_i(n-L+1)
\end{bmatrix}^T
\end{align*}
Using the part of the $\mathbf{\hat{S}}$ matrix below the second drawn line, we can estimate the missing $i$-th channel RF data $\hat{\mathbf{x}}_i^{B+1}$:
\begin{align*}
\hat{\mathbf{x}}_i^{B+1} &= 
\begin{bmatrix}
\hat{x}_i^{B+1}(0) &\hat{x}_i^{B+1}(1) \cdots  & \hat{x}_i^{B+1}(L_s-1)
\end{bmatrix}^T
\end{align*}
However there is one problem with this estimated data. Using the blind \textit{b}MCFLMS and R-MINT algorithms, we can only estimate the TRFs and PSF up to a scaling factor. As a result, we need to evaluate these scale factors of each channel to use this extra missing data block for the performance improvement of the proposed adaptive algorithm. Following a similar approach as in \eqref{eqn:data_est} and using the part of the $\mathbf{\hat{S}}$ matrix above the first drawn line, we can estimate $\hat{\mathbf{x}}_i^1$ and compare it with our received RF data from the ultrasound scanner to get the $i$-th channel scaling factor $\nu_i$ as
\begin{equation}\label{eqn:scale}
\nu_i = \frac{1}{L_b}\sum_{j=0}^{L_b-1} \frac{\tilde{\mathbf{x}}_i^1(j)}{\hat{\mathbf{x}}_i^1(j)}
\end{equation}
Using \eqref{eqn:data_est} and \eqref{eqn:scale}, an estimate of the missing RF data vector for the $i$-th channel is given by
\begin{equation}\label{eqn:missing_rf}
\tilde{\mathbf{x}}^{B+1}_i = \nu_i \hat{\mathbf{x}}^{B+1}_i
\end{equation}
With this estimated $(B+1)$-th block of the missing data we can evaluate the $(B+1)$-th block of the error function. An improved cost function is now defined as
\begin{equation}\label{eqn:split_cost}
J^{b'}(m) = \alpha_1 J^{b}(m) + \alpha_2 J^{(B+1)}(m) - \psi(m) J_{corr}^b(m)
\end{equation}
where
\begin{equation}
J^{B+1} = \sum_{i=1}^{M-1} \sum_{j=i+1}^{M}\underline{\mathbf{e}}_{ij}^{(B+1)H} \underline{\mathbf{e}}_{ij}^{(B+1)}
\end{equation}
Here $J^{(B+1)}$ contains the estimated data and hence it is given less weight in the total cost function than $J^{b}$ by suitably introducing two constants $\alpha_1$ and $\alpha_2$.
For the $(B+1)$-th block of data, \eqref{eqn:last_err} and \eqref{eqn:first_err} become
\begin{equation}\label{eqn:last_err2}
\mathbf{e}_{ij}^{p(B-p+1)} = \mathbf{C}_{x_i^p}\mathbf{h}_j^{(B-p+1)} - \mathbf{C}_{x_j^{p}}\mathbf{h}_i^{(B-p+1)},
p = 1,2,\cdots,B
\end{equation}

\begin{equation}\label{eqn:first_err2}
\mathbf{e}_{ij}^{p(B-p+2)} = \mathbf{C}_{x_i^p}\mathbf{h}_j^{(B-p+2)} - \mathbf{C}_{x_j^{p}}\mathbf{h}_i^{(B-p+2)},
p = 2,\cdots,B+1 
\end{equation}
Here in \eqref{eqn:first_err2}, $p\neq1$ as there is no $(B+1)$-th block for TRFs. Therefore, the error function $\mathbf{e}_{ij}^{B+1}$ and the cost function $J^{B+1}$ for the $(B+1)$-th block can be written as
\begin{equation}\label{eqn:err_b+1}
    \tilde{\mathbf{e}}_{ij}^{B+1} =  \sum_{p=1}^{B}A_1\mathbf{e}_{ij}^{p(B-p+1)} + \sum_{p=2}^{B+1}A_2\mathbf{e}_{ij}^{p(B-p+2)}
\end{equation}
A general form of the gradient of $\mathbf{J}^{B+1}$ with respect to $\underline{\mathbf{\hat{h}}}_{k}^{b*}$ for any block $b$ and channel $k$ can be obtained as
\begin{align}
 \frac{\partial J^{B+1}}{\partial \underline{\mathbf{\hat{h}}}_{k}^{b*}} 
                        = &\sum_{i=1}^{M}(\underline{\mathbf{x}}_i^{(B+1-b)*}.*\mathbf{B}_1^H\tilde{\underline{\mathbf{e}}}_{ik}^{(B+1)} +\nonumber \\
                        &\underline{\mathbf{x}}_i^{(B+2-b)*}.*\mathbf{B}^H\tilde{\underline{\mathbf{e}}}_{ik}^{(B+1)})
\end{align}
Now, the gradient of \eqref{eqn:split_cost} is given by
\begin{align}\label{eqn:J^b'}
\nabla_b J^{b'}(m) =& \alpha_1\nabla_b J^{b}(m) + \alpha_2\nabla_b J^{B+1}(m) \nonumber\\
 							 &- \psi(m) \nabla_b J_{corr}^b(m)
\end{align}
General expressions for $\nabla_b J^{b}(m)$ and $\nabla_b J_{corr}^b(m)$ are shown in \eqref{eqn:J^bt}.\\
The parameter update equation for the missing data estimation based \textit{b}MCFLMS (\textit{md-b}MCFLMS) algorithm for the $b$-th block of data is given by 
\begin{equation}\label{eqn:update_eqn2}
\hat{\underline{\mathbf{h}}}^b(m+1) = \hat{\underline{\mathbf{h}}}^b(m) - \mu^b(m) \nabla_b J^{b'}(m)|_{\underline{\mathbf{h}} = \hat{\underline{\mathbf{h}}}(m)}
\end{equation}
where,
\begin{align}
&\nabla_b J^{b'}(m) = \frac{\partial J^{b'}(m)}{\partial \underline{\hat{\mathbf{h}}}^b(m)}\nonumber\\
&=
\begin{bmatrix}
\frac{\partial J^{b'}(m)}{\partial \underline{\hat{\mathbf{h}}}_{1}^b(m)} & \frac{\partial J^{b'}(m)}{\partial \underline{\hat{\mathbf{h}}}_{2}^b(m)}  & .........  & \frac{\partial J^{b'}(m)}{\partial \underline{\hat{\mathbf{h}}}_{M}^b(m)}
\end{bmatrix}
\end{align}
\begin{align}\label{eqn:mu_2}
\mu^b(m) = \frac{\hat{\underline{\mathbf{h}}}^{T}(m)}{||\nabla J^{b'}(m)||^2}\nabla J^{b'}(m)
\end{align}
Here,\\
$\underline{\hat{\mathbf{h}}}^b(m)$ =
$
\begin{bmatrix}
\underline{\hat{\mathbf{h}}}_{1}^b(m) & \underline{\hat{\mathbf{h}}}_{2}^b(m)  & .........  & \underline{\hat{\mathbf{h}}}_{M}^b(m)
\end{bmatrix}
$\\
\begin{align}\label{eqn:grad}
\nabla J^{b'}(m) &= \frac{\partial J^{b'}(m)}{\partial \hat{\underline{\mathbf{h}}}(m)} \nonumber\\
&=
\begin{bmatrix}
(\frac{\partial J^{b'}(m)}{\partial \underline{\hat{\mathbf{h}}}^1(m)})^T & (\frac{\partial J^{b'}(m)}{\partial \underline{\hat{\mathbf{h}}}^2(m)})^T & \cdots  & (\frac{\partial J^{b'}(m)}{\partial \underline{\hat{\mathbf{h}}}^b(m)})^T
\end{bmatrix}^T
\end{align}
and 
\begin{align}\label{eqn:TRF}
\underline{\hat{\mathbf{h}}}(m) =
\begin{bmatrix}
\underline{\hat{\mathbf{h}}}^{1T}(m) & \underline{\hat{\mathbf{h}}}^{2T}(m) & ......... & \underline{\hat{\mathbf{h}}}^{BT}(m)
\end{bmatrix}^T
\end{align}

As before, the TRFs after each update are normalized to avoid the trivial zero solution, i.e.,
\begin{equation}\label{eqn:normalize2}
\underline{\hat{\mathbf{h}}}(m+1) = \frac{\underline{\hat{\mathbf{h}}}(m+1)}{||\underline{\hat{\mathbf{h}}}(m+1)||}
\end{equation}
\noindent
A summary of the \textit{md-b}MCFLMS algorithm is presented in Table II.

\begin{table}[h!]  \label{tab: m-mcflms}
  %  \centering
  \caption{Constrained \textit{md}-\textit{b}MCFLMS Algorithm}
  \rule{8.5cm}{0.4pt}
  \begin{enumerate}
  \item[\bf{Step 1}]
  \begin{enumerate}
  \item[.] Set appropriate value for $\psi$, $\rho$ and $\gamma$
  \item[.] Initialize the $i$-th channel TRF, $\mathbf{\hat{h}}_i $ with the TRFs estimated using \textit{b}MCFLMS algorithm
   \end{enumerate}
   
   \item[\bf{Step 2}]
    \begin{enumerate}
    \item[.] Estimate the PSF $\hat{\mathbf{s}}$ using \eqref{eqn:g_r}, \eqref{eqn:s_r} and \eqref{eqn:hat_h}
    \end{enumerate}

    \item[\bf{Step 3}]
    \begin{enumerate}
    \item[.] Estimate the scaled missing data using \eqref{eqn:data_est}
    \item[.] Using \eqref{eqn:scale} and \eqref{eqn:missing_rf} calculate the missing RF data
    \end{enumerate}
        
\item[\bf{Step 3}]
    \begin{enumerate}
    \item[.] Set current block number $b = 1$
    \end{enumerate}
    
\item[\bf{Step 4}]
    \begin{enumerate}
    \item[.] Set iteration index $m = 1$
    \end{enumerate}
 
 \item[\bf{Step 5}]
    \begin{enumerate}
    \item[.] Calculate the error functions for $b$-th block using \eqref{eqn:last_err}, \eqref{eqn:first_err}, \eqref{eqn:error}, \eqref{eqn:err_hb} and \eqref{eqn:err_b+1}
    \item[.] Calculate correlation of estimated TRF with RF data using \eqref{eqn:correlation}
    \end{enumerate}
        
\item[\bf{Step 7}]
    \begin{enumerate}
    \item[.] Calculate $\frac{\partial J^{b'}}{\partial \underline{\mathbf{\hat{h}}}_{k}^{b*}}$ according to \eqref{eqn:J^b'}
    \item[.] Calculate step-size for $b$-th block and $m$-th iteration $\mu^b(m)$ using \eqref{eqn:mu_2},  \eqref{eqn:grad} and \eqref{eqn:TRF}
    \end{enumerate}  
       
\item[\bf{Step 8}]
    \begin{enumerate}
    \item[.] Update $\underline{\mathbf{\hat{h}}}$ using \eqref{eqn:update_eqn2}
    \item[.] Normalize $\underline{\mathbf{\hat{h}}}$ according to \eqref{eqn:normalize2}
    \end{enumerate}
    
\item[\bf{Step 9}]
    \begin{enumerate}
    \item[.] If m is less than required iterations, set $m = m + 1$ and go to step $5$
    \item[.] Else set $b = b + 1$, $m = 1$ and go to step $4$
    \end{enumerate}    
  \end{enumerate}
  \rule{8.5cm}{0.4pt}
  \end{table}
%%%%%%%%%%%%%%%%%%%%%%%%%%%%%%%%%%%%%%%%%%%%
\section{RESULTS}
\begin{figure*}[ht!]
\centering
\includegraphics[width = 7 in, height = 4.3 in]{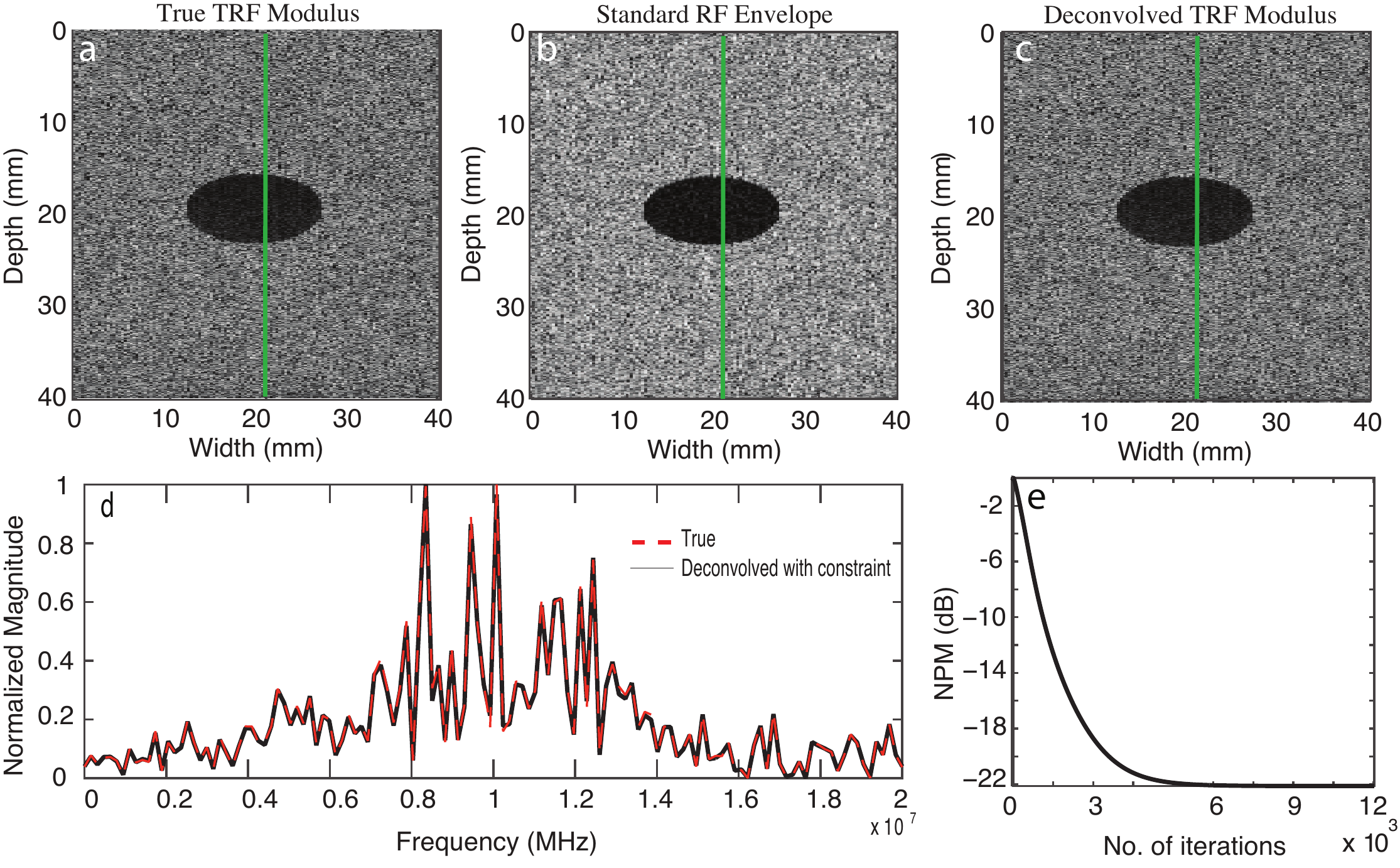}
 \caption{Deconvolution performed on a simulation phantom with $30$ scatterers per resolution cell and SNR = $30$ dB. The RF data size is $1024\times128$. The darker oval inclusion with long axis $5$ mm and short axis $2$ mm is created by placing scatterers with relatively lower strength than the surroundings. Log-envelope image of the (a) true TRF, (b) backscattered standard RF data, and (c) deconvolved TRF by the \textit{md-b}MCFLMS algorithm, (d)  spectra of the true and estimated RF data of a single scan-line marked by vertical green line in the log envelope images of (b) and (c). (e) The NPM curve between the deconvolved TRF and true TRF of the first block.}
 \label{phantom_performance}
\end{figure*}

\begin{figure*}[ht!]
\centering
\includegraphics[width = 7 in, height = 5.1 in]{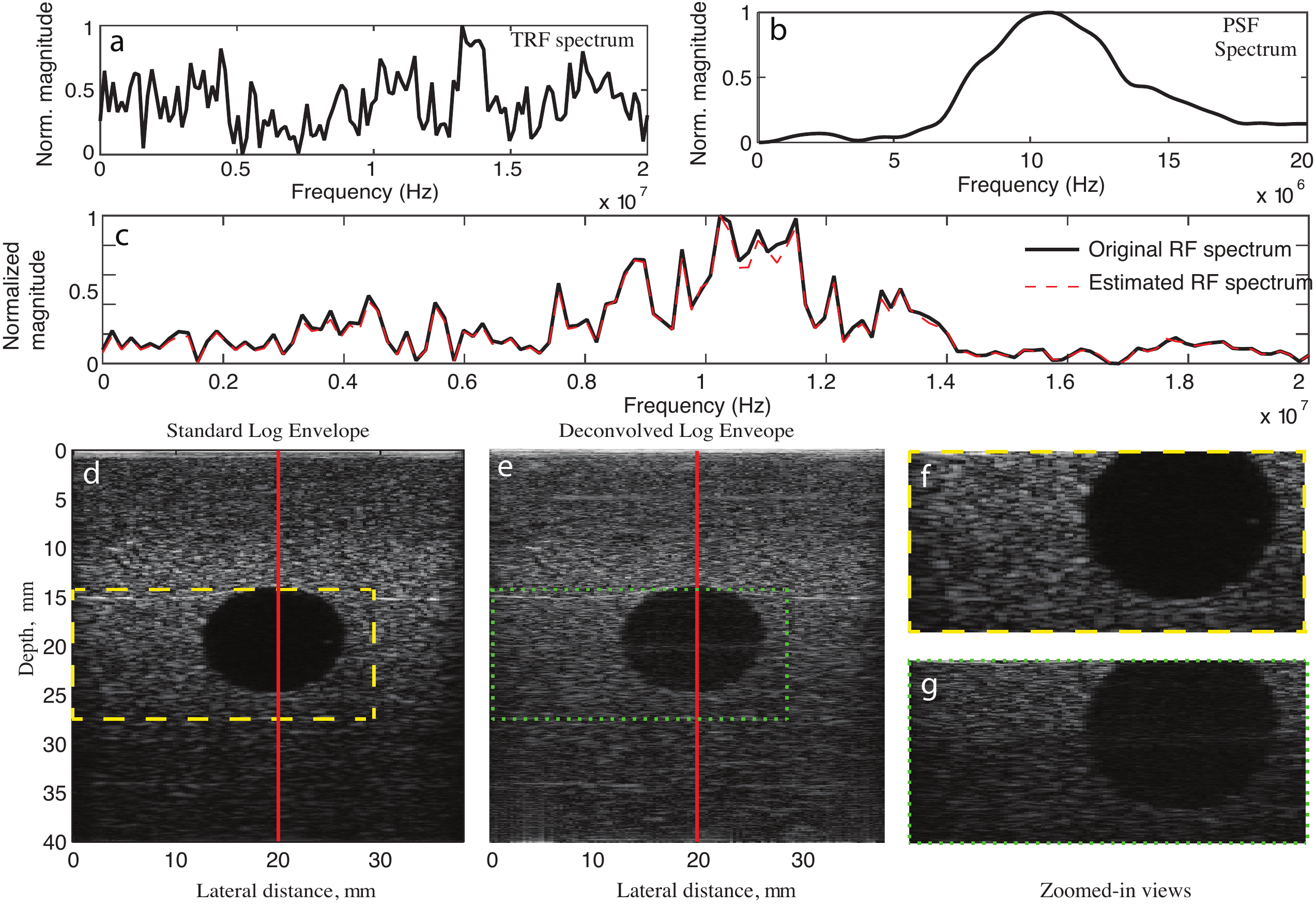}
\caption{Performance analysis of the \textit{md-b}MCFLMS algorithm using a CIRS phantom. Spectrum of the (a) deconvolved TRF, (b) R-MINT estimated PSF, and (c) spectra of the true and estimated RF data of a single scan-line marked by vertical red line in log envelope images. Standard log envelope images of the (d) backscattered RF data, and (e) deconvolved TRFs, and (f-g) zoomed-in views of (d-e).}
 \label{cirs_performance}
\end{figure*}

\begin{figure*}[ht!]
\centering
\includegraphics[width = 7 in, height = 5.1 in]{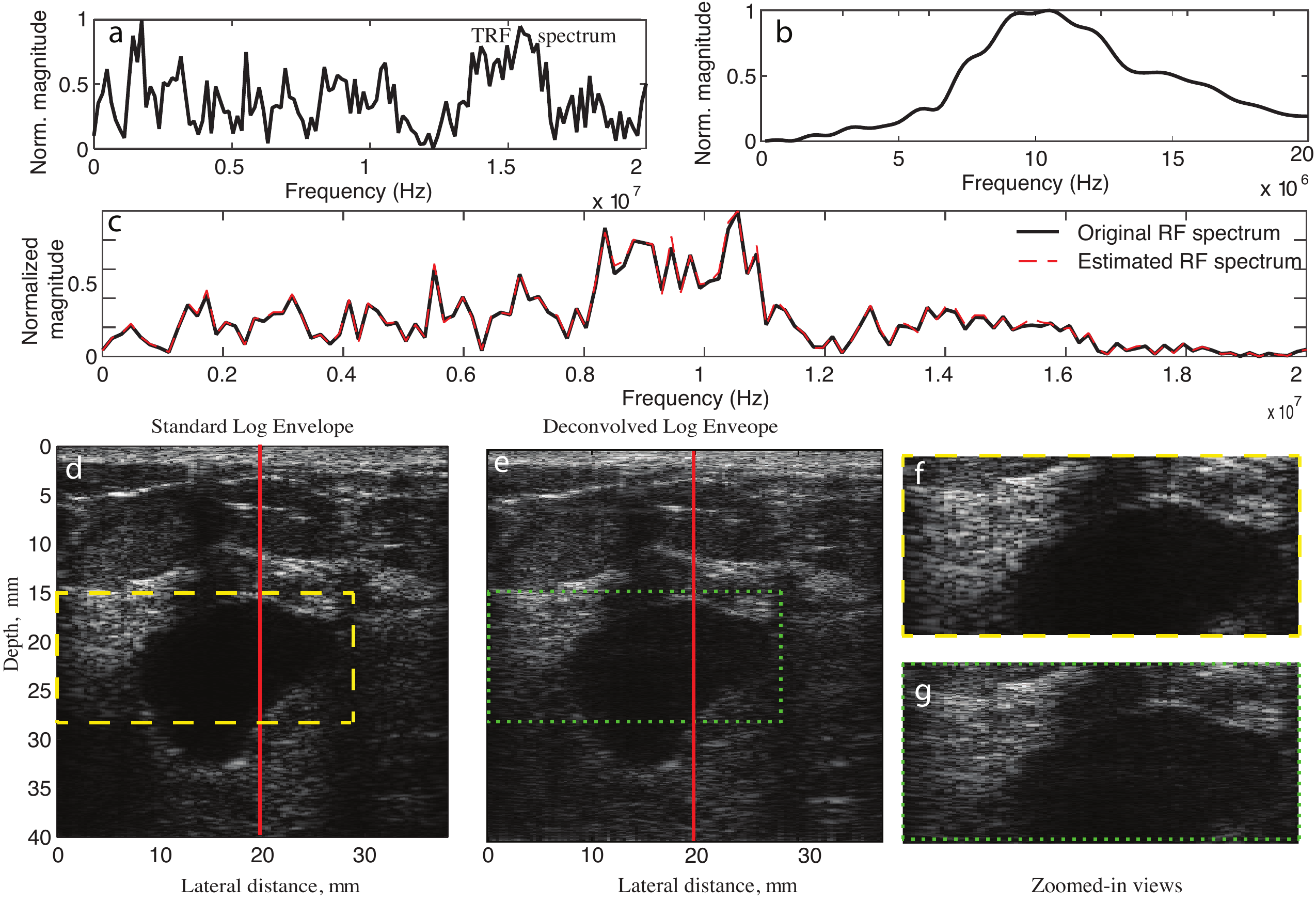}
 \caption{Performance analysis of the \textit{md-b}MCFLMS algorithm for the \emph{in-vivo} backscattered RF data of a breast cyst. Spectrum of the (a) deconvolved TRF, (b) R-MINT estimated PSF, and (c) spectra of the true and estimated RF data of a single scan-line marked by vertical red line in log envelope images. Standard log envelope images of (d) the backscattered RF data, (e) deconvolved TRFs, and (f-g) zoomed-in views of (d-e).}
 \label{cyst_performance}
\end{figure*}

\begin{figure*}[ht!]
\centering
\includegraphics[width = 7 in, height = 5.1 in]{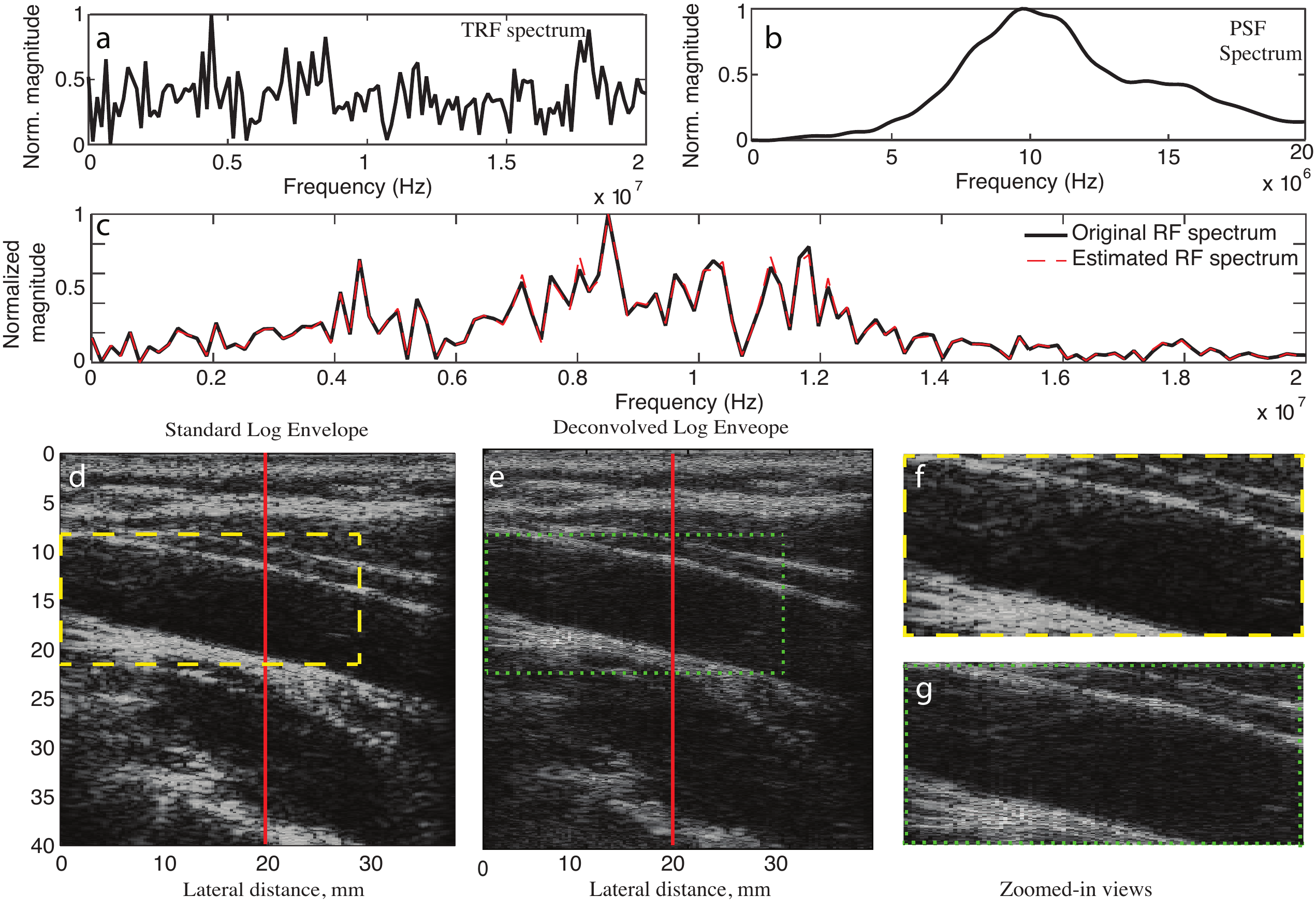}
 \caption{Performance analysis of the \textit{md-b}MCFLMS algorithm for the \emph{in-vivo} backscattered RF data for a left carotid artery. Spectrum of the (a) deconvolved TRF, (b) R-MINT estimated PSF, and (c) spectra of the true and estimated RF data of a single scan-line marked by vertical red line in log envelope images. Standard log envelope images of (d) the backscattered RF data, (e) deconvolved TRFs, and (f-g) zoomed-in views of (d-e).}
 \label{carotid_performance}
\end{figure*}

In this section, the performance of the proposed algorithms is measured on simulation phantom, experimental phantom and also on \emph{in-vivo} RF data. The results obtained are compared with the time-domain method described in \cite{hasan2016blind}, cepstrum method \cite{taxt1995restoration} and CR based method \cite{yu2012blind}. Here the performance is measured using two indices, one is the normalized projection misalignment (NPM) and the other one is the resolution gain (RG) \cite{abeyratne1995higher}. The NPM is defined as 
\begin{align}\label{eqn:NPM}
    \mbox{NPM}(m) &= 20log_{10}\left(\frac{\|\zeta(m)\|}{\|\mathbf{h}\|}\right) \\
    \zeta(m) &= \mathbf{h}-\frac{\mathbf{h}^T\mathbf{\hat{h}}(m)}{\mathbf{\hat{h}}^T(m)\mathbf{\hat{h}}(m)}\mathbf{\hat{h}}(m)
\end{align}
where $\mathbf{h}$ and $\mathbf{\hat{h}}(m)$ represent the true and estimated TRFs, respectively. Measurement of NPM requires the true TRF and hence it can be calculated only for the simulation phantom data where the true TRF is known. The RG is defined as 
\begin{equation}\label{eqn:RG_two_level}
    G_d = \frac{R^o_d}{R^d_d}, d=5dB, 10dB
\end{equation}
where $R^o_d$ and $R^d_d$ represent resolutions before and after the deconvolution, respectively.  To calculate $R^o_d$ and $R^d_d$, the normalized $2$-D autocovariance function of the RF and the TRF data are calculated. Then the axial slice through the peak is considered and the width of the slice at a level $d$ dB is measured which represents $R^d_d$.

In all the subsequent figures the reference images were generated from the RF data which have the blurring effect in it introduced by the PSF. To produce the reference images absolute value of the Hilbert transform of the RF data was taken. Then as stated in \cite{taxt1995restoration} a log compression was performed using $\log(cx + 1)$ where $c$ was adjusted to match the contrast between the standard reference and deconvolved images. Here for all data the total number of block was set at $B = 2$ and the parameters of noise effect compensating constraint were set to $\xi = 1e-4$, $\rho = 2.55$ and $\gamma = 2.4$. Again for the \textit{md}-bMCFLMS algorithm the selected parameters are $\alpha_1 = 0.1$ and  $\alpha_2 = 2.7\times10^{-5}$. The reason behind significantly small value of $\alpha_2$ is that in \eqref{eqn:split_cost} $J^b$ portion was minimized in the first step, i.e., in \textit{b}MCFLMS and to make $J^{(B+1)}$ comparable to $J^b$, $J^{(B+1)}$ is given a smaller weight.

\subsection{Simulation Phantom Results}
The ultrasound simulation was done using the FIELD-II \cite{jensen2004simulation} where the transducer element height was chosen to be 5 mm. Ultrasound simulation was done on a 3-D simulation phantom ($40$ mm $\times 10$ mm $\times 40$ mm) with a scatterer density of 30 scatterers per resolution cell. An oval inclusion with long axis $5$~mm and short axis $2$~mm was simulated by reducing the magnitudes in that region. The focus of the ultrasound beam was set at 30 mm depth from the phantom surface. The transducer centre frequency was selected as 10 MHz and the sampling frequency as 40~MHz. An array of 128 transducer elements was used to match with that available in commercial ultrasound scanners. To simulate noisy data zero-mean additive white Gaussian noise was added to the data so as to obtain an SNR of $30$ dB.

The performance of the proposed \textit{md-b}MCFLMS algorithm on the simulation data is shown in Fig. \ref{phantom_performance}. From visual comparison of the images provided in Figs. \ref{phantom_performance}(a)-(c) we see that the standard RF image (Fig. \ref{phantom_performance}(b)) has poor resolution in contrast to our estimated TRF (Fig. \ref{phantom_performance}(c)) that matches closely in terms of resolution with the true TRF image (Fig. \ref{phantom_performance}(a)). Fig. \ref{phantom_performance}(d) shows that the estimated TRF along the green marked line of the image matches closely with that of the true TRF. Again,  since we have the true TRF for the simulation data, we can evaluate NPM for our estimated TRF which is shown in Fig. \ref{phantom_performance}(e). Fig. \ref{phantom_performance}(e) proves the effectiveness of our proposed correlation-based constraint to prevent misconvergence of the proposed algorithm as the NPM remains stable at $-22.1$ dB. 

Comparative results of different algorithms on simulation phantom data at SNR $= 30$ dB are presented in Table \ref{tab: table1}. The quantitative performance measures used for comparison are NPM (see \eqref{eqn:NPM}) and $G_d$ (see \eqref{eqn:RG_two_level}). It is obvious from Table \ref{tab: table1} that our proposed \textit{b}MCFLMS algorithm gives better image quality in terms of resolution gain ($5$ and $10$ dB level) and NPM than that of the other methods. In order to show the improvement after using the missing data in our algorithm, results on both the initial estimate from the \textit{b}MCFLMS and final deconvolved image  from the \textit{md-b}MCFLMS are presented. Table \ref{tab: table1} shows that the \textit{md-b}MCFLMS algorithm improves the RG and NPM compared to the \textit{b}MCFLMS proposed in this paper. It requires additional $127$ iterations as compared to the \textit{b}MCFLMS algorithm to improve the image further.

\begin{table}[h!]
\caption{Performance of different algorithms on simulation phantom data with noise}
\label{tab: table1}
\resizebox{\columnwidth}{!}{
\begin{tabular}{||c|c||c|c||c|}\hline
\multicolumn{1}{||c|}{ } &
\multicolumn{1}{c||}{ } &
\multicolumn{1}{c|}{ } &
\multicolumn{2}{c|}{$G_d$(RF)} \\ \cline{4-5}
Data & Method & NPM (dB) & 5 dB & 10 dB \\ \hline
Simulation data & CR-based Method  & -15.125 & 3.824 & 4.2321\\ \cline{2-5}
				 & Cepstrum 				 & -16.105 & 4.1055 & 5.1023\\ \cline{2-5}
				 & \textit{$l_1$-b}MCLMS 		 & -19.9974 & 4.2097 & 5.1376\\ \cline{2-5}
				 & Proposed \textit{b}MCFLMS  & -21.802 & 4.5189 & 6.0293\\ \cline{2-5}
				 & Proposed \textit{md-b}MCFLMS   & -22.105 & 4.7213 & 6.214\\ \hline
\end{tabular}
}
\end{table}

\subsection{Experimental Phantom Results}
The performance of our proposed algorithms (i.e., \textit{b}MCFLMS and \textit{md-b}MCFLMS) was also tested on RF data generated from an experiment on a CIRS (Computerized Imaging Reference System, Norfolk, Virginia, USA) tissue mimicking phantom (TMP) of dimension $18 \times 12 \times 9.5$~cm. It has a background made of zerdine with an attenuation coefficient of $0.70$ db/cm/MHz and a sound velocity of $1540$~m/s. It has a spherical inclusion of radius $6.8$ mm with an attenuation coefficient of 1.05 dB/cm/MHz. The backscattered data from the TMP was collected using a SonixTOUCH Reasearch (Ultrasonix Medical Corp., Richmond BC, Canada) with a L14-5/38 probe operating at 10 MHz and sampling frequency of 40 MHz.

Fig. \ref{cirs_performance} shows the performance of our proposed \textit{md-b}MCFLMS algorithm on the experimental phantom data. Fig. \ref{cirs_performance}(a) shows the wideband TRF spectrum where the TRF line is chosen along the red line of Fig. \ref{cirs_performance}(e) and Fig. \ref{cirs_performance}(b) shows the narrowband, smooth PSF spectrum. The estimated RF spectrum evaluted from the multiplication of TRF spectrum and PSF spectrum, matches closely with true RF spectrum which is shown in Fig. \ref{cirs_performance}(c). This ensures the deconvolution efficacy of the proposed algorithm. From Figs. \ref{cirs_performance}(d)-(e), we get a visual proof of resolution increase in the deconvolved image compared to that of the standard RF image. For better visualization, zoomed-in view of a segment of Figs. \ref{cirs_performance}(d)-(e) are given in Figs. \ref{cirs_performance}(f)-(g). As can be seen, the deconvolved image has finer texture compared to the standard RF image.

The quantitative performance results in terms of resolution gain presented in Table \ref{tab: table2} show that the time-domain \textit{$l_1$-b}MCLMS algorithm performs better than the CR-based and cepstrum methods. However, our proposed \textit{b}MCFLMS algorithm in the frequency-domain gives better resolution gain at $5$ and $10$ dB level than that of the \textit{$l_1$-b}MCLMS algorithm and it is further improved by the proposed \textit{md-b}MCFLMS algorithm.
\begin{table}[h!]
\caption{Performance of different algorithms on experimental phantom and \emph{in-vivo} data}
\label{tab: table2}
\resizebox{\columnwidth}{!}{
\begin{tabular}{||c|c||c|c|}\hline
\multicolumn{1}{||c|}{ } &
\multicolumn{1}{c||}{ } &
\multicolumn{2}{c|}{$G_d$(RF)}\\ \cline{3-4}
Data & Method & 5 dB & 10 dB\\ \hline
CIRS Phantom & CR-based Method & 3.42 & 2.22\\ \cline{2-4}
				 & Cepstrum 				& 1.98 & 2.41\\ \cline{2-4}
				 & $l_1$-\textit{b}MCLMS 			& 3.86 & 4.02\\ \cline{2-4}
				 & Proposed \textit{b}MCFLMS 			& 4.79 & 6.63\\ \cline{2-4}
				 & Proposed \textit{md-b}MCFLMS  & 4.88 & 6.82\\ \hline \hline
Breast Cyst & CR-based Method & 2.25 & 0.9585\\ \cline{2-4}
				 & Cepstrum 				& 2.38 & 1.12\\ \cline{2-4}
				 & \textit{$l_1$-b}MCLMS 			& 3.16 & 4.64\\ \cline{2-4}
				 & Proposed \textit{b}MCFLMS 			& 3.58 & 6.17\\ \cline{2-4}
				 & Proposed \textit{md-b}MCFLMS & 3.74 & 7.02\\  \hline \hline
Carotid Artery & CR-based Method & 3.44 & 5.03\\ \cline{2-4}
				 & Cepstrum 				& 2.44 & 2.16\\ \cline{2-4}
				 & \textit{$l_1$-b}MCLMS 			& 3.72 & 5.89\\ \cline{2-4}
				 &Proposed \textit{b}MCFLMS			& 4.01 & 7.05\\ \cline{2-4}
				 & Proposed \textit{md-b}MCFLMS & 4.21 & 7.34\\  \hline
\end{tabular}
}
\end{table}

\subsection{\emph{In-Vivo} Results}
The performance of the proposed algorithms was tested on two \emph{in-vivo} data. One of these data is of breast containing a cyst and the other is of a carotid artery. These data were collected from the patients who appeared for medical examination at the Medical Centre of Bangladesh University of Engineering and Technology (BUET), Dhaka, Bangladesh. This study was approved by the Institutional Review Board (IRB) and prior patient consent was taken. 

The effect of deconvolution is more prominent in case of \emph{in-vivo} data than those of the other data mentioned earlier. This may be due to the fact that the scatterer density is more in case of \emph{in-vivo} data \cite{jirik2008two}. Fig. \ref{cyst_performance} shows the performance of our proposed \textit{md-b}MCFLMS algorithm on the breast cyst data. Figs. \ref{cyst_performance}(a)-(b) show the frequency spectrum of the TRF data for the axial line marked red in Fig. \ref{cyst_performance}(e) and the PSF. The frequency spectrum for the RF data was estimated from the product of PSF and TRF spectrum and it was superimposed on the real RF spectrum in Fig. \ref{cyst_performance}(c). The close match between the two spectrum ensures the convergence of the algorithm. From visual perspective, Fig \ref{cyst_performance}(e) shows a significant increase in resolution compared to Fig. \ref{cyst_performance}(d). For better visualization zoomed-in view of the two images are provided in Figs. \ref{cyst_performance}(f)-(g) which also show the superiority of our proposed algorithm. It is evident from Fig \ref{cyst_performance}(d)-(e) that the deconvolved image offers finer texture compared to the blurred RF image.

The results for carotid artery depicted in Fig. \ref{carotid_performance} also demonstrate similar improvements in performance by the proposed methods to that presented in Fig. \ref{cyst_performance}.

\begin{figure}[ht!]
\includegraphics[width = 3.5 in, height = 2.3 in]{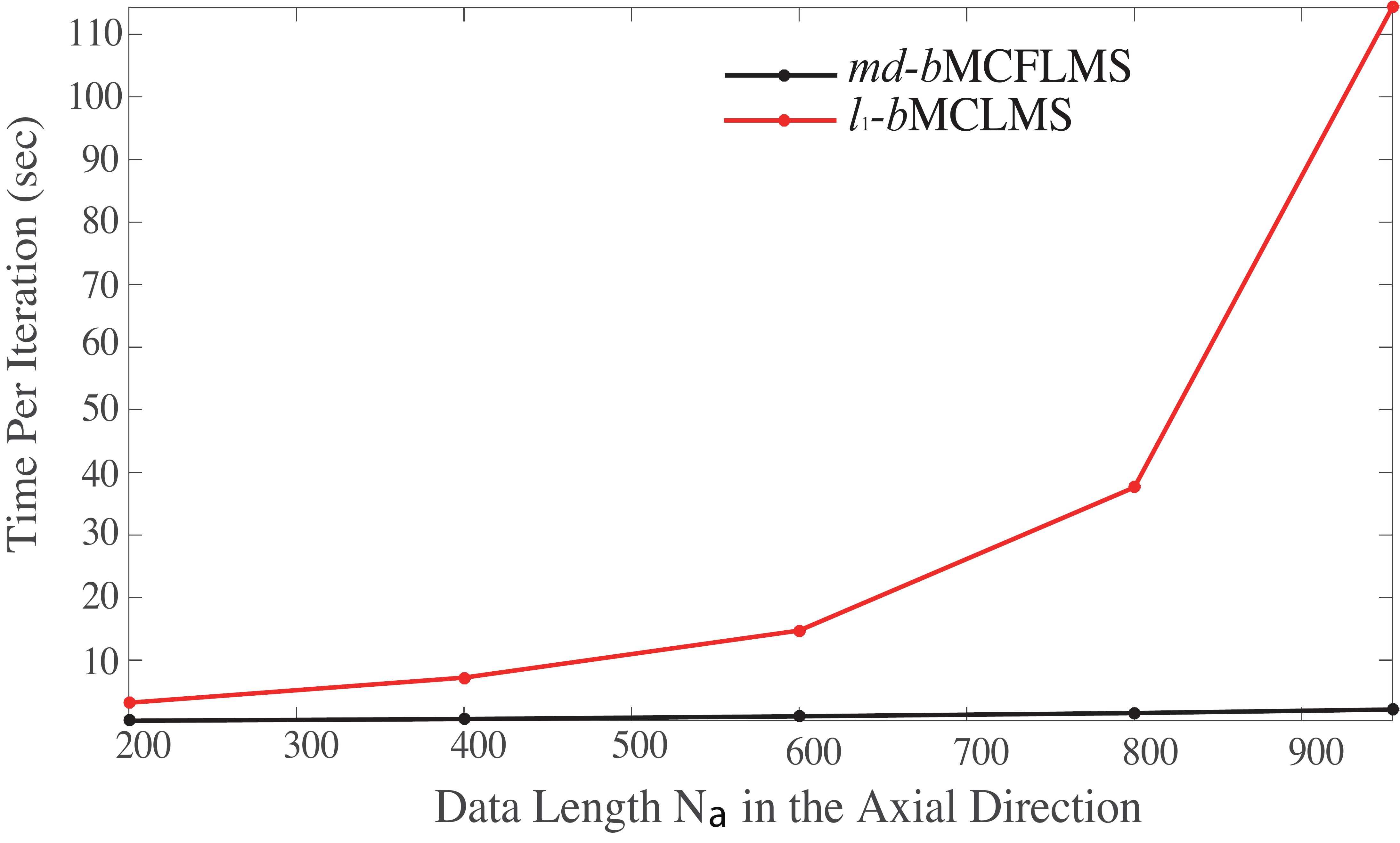}
\caption{Time required per iteration by the time-  and the frequency-domain algorithms  for the experimental phantom data of size $N_a\times128$ where $N_a$ is made variable.}
 \label{time}
\end{figure}
As shown in Table \ref{tab: table2}, our proposed \textit{b}MCFLMS algorithm gives significantly better resolution gain at $10$ dB level compared to its time-domain counterpart, CR-based method and cepstrum method for the \emph{in-vivo} data. The latter methods also show data dependency as they give resolution gain improvement in different range for the breast cyst and the carotid artery data. As discussed in the introduction section, the effectiveness of cepstrum method depends on the separability of the TRF and the PSF spectrum. Again, the CR-based method requires thresholding to find the null space bases of the correlation matrix. However, $l_1$-\textit{b}MCLMS and our proposed \textit{md-b}MCFLMS algorithm do not impose any stringent requirements on the data. Therefore, they show similar resolution improvement for both type of data which justifies their less dependence on parameter tuning. The parameters for correlation constraint once set remain effective for all types of data presented in the result section.

%%%%%%%%%%%%%%%%%%%%%%%%%%%%%%%%%%%%%%%%%%%%%%%%%%%%%%%%%%

\section{Discussion And Conclusion}
In this paper, a correlation constrained missing-data estimation based blind multichannel frequency-domain LMS-type adaptive algorithm has been proposed for ultrasonic TRF estimation. In order to exploit the inherent advantages of noise robustness and faster convergence with less number of iterations, the deconvolution has been performed in the frequency-domain. In this algorithm, to estimate the missing data, at first the TRFs are estimated using the \textit{b}MCFLMS algorithm derived in this paper and then from these estimated TRFs, the PSF is estimated using the R-MINT algorithm. Now, incorporating this additional information about the RF data, a modified cost function has been proposed in the \textit{md-b}MCFLMS algorithm that causes further improvement in the deconvolution result.  Again, to address the issue of non-stationarity of the PSF, unlike the convolution matrix segmentation described in the \textit{b}MCLMS algorithm, a time-efficient blocking procedure has been introduced in this paper. Moreover, a well-known phenomenon of misconvergence in the blind multi-channel identification algorithms in the presence of additive noise and estimation error has been addressed here. To prevent our proposed algorithm from misconverging, a novel constraint which exploits the correlation between the RF data and the estimated TRF has been proposed in this paper. The salient feature of this constraint is that it is more generalized to prevent misconvergence of the crossrelation-based blind SIMO model identification schemes from noisy measurements. This constraint also imposes maximum correlation between the RF data and the estimated TRF which causes a little more improvement in the resolution of the deconvolved ultrasound images compared to that of the misconvergence point. The efficacy of the proposed algorithm has been tested both quantitatively and qualitatively on simulation, experimental and \emph{in-vivo} data. The results have demonstrated the superiority of our proposed algorithms (\textit{b}MCFLMS and \textit{md-b}MCFLMS) compared to CR-based, cepstrum and time-domain \textit{b}MCLMS methods.

In order to show the time requirement per iteration of the proposed algorithm, the experimental phantom data described in the result section was used. The data length along the axial direction was varied while the channel number was kept fixed at $128$. The implementation platform used were: CPU: Intel$\textsuperscript{\textregistered}$ Core$\textsuperscript{TM}$ i5, RAM: 8 GB, software:
MATLAB$\textsuperscript{\textregistered}$, The MathWorks, Natick, MA. The plot in Fig. \ref{time} shows that our proposed algorithm takes significantly less time per iteration than is required by the time-domain \textit{b}MCLMS algorithm. In addition, because of reduced eigen-value spread, it is observed that the frequency-domain implementation requires fewer iterations for convergence than the time-domain approach. This implies that our proposed algorithm is faster compared to the \textit{b}MCLMS algorithm. The computation time of the algorithm may be further reduced by implementing it in JAVA or C.

There is a scope for improving the quality of the TRF image further by considering a more realistic $2$-D model for the PSF instead of its $1$-D approximation that is used here. Again, the coupling factor used with the correlation constraint has been chosen empirically and an elegant solution for this issue will be addressed in our future work. Nevertheless, our proposed algorithm results in a better image quality of ultrasound images by removing the PSF effect from the RF data which may have far-reaching effect on tissue characterization using quantitative ultrasound.

%\ifCLASSOPTIONcaptionsoff
 % \newpage
%\fi
\ifCLASSOPTIONcaptionsoff
  \newpage
\fi
\section*{Acknowledgment}
This work has been supported by the Higher Education Quality Enhancement Program (HEQEP) of University Grants Commission (UGC), Bangladesh (CPSF-96/BUET/W-2/2017).
\ifCLASSOPTIONcaptionsoff
  \newpage
\fi
\bibliographystyle{IEEEtran}

\bibliographystyle{IEEEtran}
\bibliography{jd}

\end{document}